\documentclass[rmp,aps,twocolumn,floatfix,unsortedaddress,reprint]{revtex4-1}

\usepackage{graphics}
\usepackage{epsfig}
\usepackage{amsmath}

\usepackage{amsmath,amssymb,bm,curves}

\usepackage{lineno}

\def\inbar{\,\vrule height1.5ex width.4pt depth0pt}
\def\IR{\relax{\rm I\kern-.18em R}}
\def\IC{\relax\hbox{$\inbar\kern-.3em{\rm C}$}}

\def\ubar{{\bar{u}}}
\def\dbar{{\bar{d}}}
\def\sbar{{\bar{s}}}

\newcommand{\be}{\begin{linenomath}\begin{equation}}
\newcommand{\ee}{\end{equation}\end{linenomath}}
\newcommand{\beqa}{\begin{linenomath}\begin{eqnarray}}
\newcommand{\eeqa}{\end{eqnarray}\end{linenomath}}

\newcommand{\footfrac}[2]%

\newcommand{\open}{{<\kern -0.3em{\scriptscriptstyle )}}}

\begin{document}


\title{Etaprime and Eta Mesons with Connection 
to Anomalous Glue}

\author{Steven D. Bass}
\email{Steven.Bass@cern.ch}
\affiliation{\mbox{Kitzb\"uhel Centre for Physics,
Kitzb\"uhel, Austria}}
\affiliation{\mbox{Marian Smoluchowski Institute of Physics, Jagiellonian University, 
PL 30-348 Krakow, Poland}}
%
\author{Pawel Moskal}
\email{P.Moskal@uj.edu.pl}
\affiliation{\mbox{Marian Smoluchowski Institute of Physics, Jagiellonian University, 
PL 30-348 Krakow, Poland}}

\begin{abstract}
We review the present understanding of 
$\eta'$ and $\eta$ meson physics and these mesons as a 
probe of gluon dynamics in low-energy QCD.
Recent highlights include the production mechanism of
$\eta$ and $\eta'$ mesons in proton-nucleon collisions
from threshold to high-energy,
the $\eta'$ effective mass shift in the nuclear medium,
searches for possible $\eta$ and $\eta'$ bound states
in nuclei as well as precision measurements of $\eta$ 
decays as a probe of light-quark masses.
We discuss recent experimental data, theoretical interpretation 
of the different measurements and the open questions and challenges for future investigation.
\end{abstract}

\date{1 October 2018}
\maketitle
\tableofcontents

\section{Introduction}

The $\eta'$ meson is special in Quantum Chromodynamics, 
the theory of quarks and gluons (QCD), 
because of its strong affinity to gluons.
Hadrons, their properties and interactions, are emergent 
from more fundamental QCD quark and gluon degrees of freedom. 
QCD has the property of asymptotic freedom. 
The coupling $\alpha_s (P^2)$ which describes the strength of quark-gluon 
and gluon-gluon interactions
decreases logarithmically with increasing 
(large) four-momentum transfer squared, $P^2$.
In the infrared, at low $P^2$,
quark-gluon interactions become strong.
Quarks become confined inside hadron bound states
and the vacuum is 
not empty but 
characterized by 
the formation of quark and gluon condensates.
The physical degrees of freedom are emergent hadrons 
(protons, mesons ...) as bound states of quarks and gluons. 
Baryons like the proton are bound states of three valence 
quarks. Mesons are bound states of a quark and antiquark.

Glue is manifest in the confinement potential which 
binds the quarks.
This confinement potential corresponds to a 
restoring force of 10 tonnes regardless of separation.
Quarks are bound by a string of glue 
which can break into two colorless hadron objects
involving the creation of a quark-antiquark 
pair corresponding to the newly created ends of 
two confining strings formed from the original
single string of confining glue.
There are no isolated quarks.
The QCD confinement radius is of order 1fm=$10^{-15}$m.
This physics at large coupling 
is beyond QCD perturbation theory and described either
using QCD inspired models of hadrons
which build in key symmetries of the underlying theory
or through computational lattice methods.
About 99\% of the mass of the 
hydrogen atom, 938.8 MeV, 
is associated with the confinement potential
with the masses of 
the electron 0.5 MeV and
the proton 938.3 MeV.
Inside the proton 
the masses of the proton's constituent 
two up quarks and one down quark are about 
2.2 MeV for each up quark and 4.7 MeV for the down quark.

Besides generating the QCD confinement potential,
glue plays a special role
in the light hadron spectrum 
through the physics 
of the isoscalar
$\eta'$ and $\eta$ mesons
including their interactions.
The QCD Lagrangian with massless quarks is symmetric
between left- and right-handed quarks
(which are fermions)
or between positive and negative helicity quarks.
However, this symmetry is missing in the ground state 
hadron spectrum.
The lightest mass hadrons, pions and kaons, 
are pseudoscalar mesons called Goldstone bosons 
associated with 
the spontaneous breaking of 
chiral symmetry between left- and right-handed quarks.
These mesons are special in that
the square of their masses are proportional 
to the masses of their constituent valence quark-antiquark pair.
(In contrast, the leading term in the masses of the proton
 and spin-one vector mesons is determined by the confining gluonic
 potential with contributions from the light quark masses
 treated as small perturbations.)
The lightest mass pions,
the neutral $\pi^0$ with mass 135 MeV 
and charged $\pi^{\pm}$ with mass 140 MeV,
play an important role 
in nuclear physics and the nucleon-nucleon interaction.
The isosinglet partners of the pions and kaons, 
the pseudoscalar $\eta$ and 
$\eta'$ mesons are too massive by about 300-400 MeV 
for them to be pure Goldstone states. 
They receive extra mass from non-perturbative gluon 
dynamics 
through a quantum effect called the axial anomaly.
This glue comes with non-trivial topology.
The physics of Goldstone bosons and 
the axial anomaly are explained in Section II below.
Gluon topology is an effect beyond the simplest quark
models and involves non-local and 
long range properties of the gluon fields.
Theoretical understanding of the $\eta$ and $\eta'$ 
involves subtle interplay of local symmetries and non-local
properties of QCD.
Examples of 
topology in other branches of physics include 
the Bohm-Aharanov effect and 
topological phase transitions and phases of matter
in condensed matter physics, the 2016 Nobel Prize for Physics.

The $\eta$ and $\eta'$ mesons come with rich phenomenology.
The $\eta'$ is predominantly a flavor-singlet state.
This means that its wavefunction is
approximately symmetric in the three lightest 
quark types 
(up, down and strange)
that build up light hadron spectroscopy.
These different species of quarks couple to gluons with
equal strength.
The $\eta'$ meson
has strong coupling to gluonic intermediate states 
in hadronic reactions from low through to high energies.
An example from high energy reactions is the decay 
$J/\Psi \to \eta' \gamma$.
The $J/\Psi$ is made of a 
heavy charm-anticharm quark pair with mass 3686 MeV.
Its decay to the light quark $\eta'$ 
meson plus a photon
involves the annihilation of the charm-anticharm 
quark pair into a gluonic intermediate state 
which then forms the $\eta'$ meson made of a
near symmetric superposition of light quark-antiquark pairs
(up-antiup, down-antidown and strange-antistrange).

In this article we will discuss the broad spectrum 
of processes involving the $\eta'$ that are mediated 
by gluonic intermediate states.
The last 20 years has seen a dedicated programme of
$\eta'$ and $\eta$ meson production experiments from
nucleons and nuclei close to threshold as well as in 
high energy collisions.
Studies of $\eta$ and $\eta'$ meson production and 
decay processes combine to teach us about the interface 
of glue and chiral dynamics,
the physics of Goldstone bosons, in QCD.
Measurements of $\eta$ and $\eta'$ production 
in nuclear media are sensitive to behavior of 
fundamental QCD symmetries at finite density and temperature.
In finite density nuclear media,
for example in nuclei and neutron stars, 
hadrons propagate in the presence of long range mean fields 
that are created by nuclear many body dynamics.
Interaction with the mean fields in the nucleus can 
change the hadrons' observed properties,
{\it e.g.}, their effective masses, magnetic moments and 
axial charges.
Symmetries between left- and right-handed quarks,
which are spontaneously broken in the ground state, 
are partially restored in nuclear media
with a reduced size of the quark condensate.
At (large) finite temperature 
there is an effective renormalization of the QCD coupling
which becomes reduced relative to the zero temperature theory
for the same four-momentum transfer squared.
One expects changes in hadron properties 
in the interaction region of finite temperature heavy-ion collisions.
This article surveys $\eta$ and $\eta'$ meson physics as 
a probe of QCD dynamics emphasizing recent advances from experiments and theory.

In addition to the topics discussed here,
the physics of glue in QCD features in many frontline areas
of QCD hadron physics research.
The planned electron-ion-collider (EIC) 
has an exciting programme to study the role of 
glue in nucleons and nuclei over a broad range of 
high energy kinematics
\cite{Accardi:2012qut,Deshpande:2017edt}.
The search for hadrons containing explicit gluon
degrees of freedom in their bound state wavefunctions
is a hot topic in QCD spectroscopy,
{\it e.g.}, 
possible glueball states built of two or three valence 
gluons 
and hybrids built of a quark-antiquark pair and a gluon
\cite{Klempt:2007cp}.
Gluons in the proton play an essential role 
in understanding the proton's internal spin structure
\cite{Aidala:2012mv}.
Studies of the QCD phase diagram 
\cite{BraunMunzinger:2008tz}
from high density neutron stars 
\cite{Lattimer:2015nhk}
to high temperature quark-gluon plasma
and a color-glass condensate postulated to explain 
high density gluon matter in high energy collisions
\cite{Gyulassy:2004zy}
are hot topics at the interface of nuclear and particle
physics research.
On the theoretical side, 
much effort is invested in trying to understand the detailed
dynamics which leads to the QCD confinement potential
\cite{Greensite:2011zz}.

The plan of this paper is as follows.

In Section II we introduce the key theoretical issues with 
the $\eta$ and $\eta'$ mesons and their unique place at 
the interface of chiral and non-perturbative gluon dynamics. 
Here we explain the different gluonic effects at work in 
$\eta$ and
$\eta'$ meson physics and how they are incorporated in
theoretical calculations.

Section III discusses the strong CP puzzle.
The observed matter antimatter asymmetry in 
the Universe requires
some extra source of CP violation beyond 
the quark mixing described by the Cabibbo-Kobayashi-Maskawa 
(CKM) matrix in the electroweak Standard Model.
The non-perturbative glue which generates the large $\eta'$ 
mass also has the potential to break CP symmetry in the strong 
interactions.
This effect would be manifest as a finite neutron electric 
dipole moment proportional to a new QCD parameter, 
$\theta_{\rm QCD}$,
which is experimentally constrained to be very small, 
less than $10^{-10}$.
One possible explanation for the absence of CP violation here 
involves a new light-mass pseudoscalar particle called the axion.
The axion is also a possible dark matter candidate to 
explain the ``missing mass'' in the Universe.
While no axion particle has so far been observed, these
ideas have inspired a vigorous program of ongoing experimental investigation to look for them.

Sections IV-VII focus on $\eta$ and $\eta'$ phenomenology.
In Section IV we discuss the information about QCD 
which follows from $\eta$ and $\eta'$ decay processes.
The amplitude for the 
$\eta$ meson
to three pions decay 
depends on the difference between the lightest up and down
quark masses
and
provides valuable information
about the ratio of light quark masses.
Studies of $\eta$ and $\eta'$ decays tell us about
their internal quark-gluon and spatial structure. 
In addition,
searches for rare decay processes provide valuable tests of
fundamental symmetries.

Section V discusses $\eta$ and $\eta'$ production in
near-threshold proton-nucleon collisions. 
The experimental program on $\eta$ and $\eta'$ nucleon
interactions has focused on near-threshold meson production 
in proton-nucleon collisions 
and photoproduction from proton and deuteron targets
\cite{Wilkin:2016mfn,Krusche:2014ava,Metag:2017yuh,Moskal:2002jm}.
Recent highlights include the use of polarization observables 
in photoproduction experiments to search for new excited 
nucleon resonances~\cite{Anisovich:2017pox},
measurement of the $\eta'$ nucleon scattering length 
through the final state interaction in proton-proton 
collisions 
\cite{Czerwinski:2014yot}
and measurement of the spin analyzing power to probe 
the partial waves associated with $\eta$ production 
dynamics in proton-proton collisions \cite{Adlarson:2017jtw}.

Section VI deals with the $\eta$ and $\eta'$ in QCD 
nuclear media and the formation of possible 
meson-nucleus bound states. 
Recent photoproduction experiments in Bonn have revealed 
an $\eta'$ effective mass shift in nuclear medium, 
which is about 
-40 MeV at nuclear matter density \cite{Nanova:2013fxl}.
Studies of the transparency of the nuclear medium to the 
propagating $\eta'$ allow one to make a first 
(indirect) measurement of the $\eta'$-nucleus 
optical potential.
One finds a small width of the $\eta'$ in medium 
\cite{Nanova:2012vw}
compared to the depth of the optical potential 
meaning that the $\eta'$ may be a good candidate 
for possible bound state searches in finite nuclei.

Mesic nuclei, if discovered in experiments, are a new exotic state 
of matter involving the meson being bound inside the nucleus purely 
by the strong interaction, 
without electromagnetic Coulomb effects playing a role.
Strong attractive interactions between the $\eta$ meson and
nucleons mean that both the $\eta$ and $\eta'$ 
are prime targets for mesic nuclei searches, 
with a vigorous ongoing program of experiments in both 
Europe and Japan~\cite{Metag:2017yuh}.
Searches for possible $\eta$ mesic nuclei are focused on 
helium while searches for $\eta'$ bound states are focused
on carbon and copper.

The $\eta'$ effective mass shift in nuclei of about 
-40 MeV at nuclear matter density is in excellent agreement 
with the prediction of the Quark Meson Coupling model 
\cite{Bass:2005hn} which works through coupling of 
the light up and down quarks in the meson to the $\sigma$ 
(correlated two pion) mean field inside the nucleus.
Here, the $\eta'$ experiences an effective mass shift in nuclei 
which is catalyzed by its gluonic component
\cite{Bass:2013nya}.
Without this glue, the $\eta'$ would be a strange quark state 
after SU(3) breaking with small interaction with the $\sigma$
mean field inside the nucleus.

Shifting from finite density to finite temperature,
there are also hints in 
data
from RHIC 
(the Relativistic Heavy-Ion Collider)
for possible $\eta'$ mass suppression at finite 
temperature, 
with claims of at least -200 MeV 
mass shift \cite{Csorgo:2009pa,Vertesi:2009wf}.

Section VII discusses $\eta$ and $\eta'$ production 
in high-energy hadronic scattering processes from light-quark hadrons.
The ratio of $\eta$ to $\pi$ meson production at 
high transverse momentum, $p_t$, 
in high-energy proton-nucleus and nucleus-nucleus collisions 
is observed to be independent of the target nucleus 
in relativistic heavy-ion collision data from RHIC 
at Brookhaven National Laboratory and the ALICE experiment 
at the Large Hadron Collider at CERN,
indicating a common propagation through the nuclear medium 
in these kinematics.
Interesting effects are also observed in high-energy $\eta'$
production.
The COMPASS experiment at CERN found that 
odd $L$ exotic partial waves $L^{-+}$ are strongly enhanced 
in $\eta' \pi$ relative to $\eta \pi$ exclusive production 
in collisions of 191 GeV negatively charged pions from hydrogen 
\cite{Adolph:2014rpp},
consistent with expectations \cite{Bass:2001zs}
based on gluon-mediated couplings of the $\eta'$.

In Section VIII we give conclusions and an outlook to
possible future experiments which could shed new light 
on the structure and interactions of the $\eta$ and $\eta'$.

Earlier reviews on $\eta$ and $\eta'$ 
meson physics, each with a different emphasis, are 
given in the volume edited by \textcite{Bijnens:2002zy}.
The lecture notes of 
\textcite{Shore:2007yn}
provide a theoretical overview of gluonic effects 
in $\eta'$ physics.
Axion physics is reviewed in \textcite{Kawasaki:2013ae}.
\textcite{Leutwyler:2013wna} 
discusses light-quark physics
with focus on the $\eta$ meson
and
\textcite{Kupsc:2009zzb} gives an 
overview of the analysis of $\eta$ and $\eta'$ meson decays.
Meson production in proton-proton collisions 
close-to-threshold is discussed in detail in the reviews
by
\textcite{Moskal:2002jm},
\textcite{Krusche:2014ava}
and
\textcite{Wilkin:2016mfn}.
The present status of 
meson-nucleus interaction studies
is reviewed in \textcite{Metag:2017yuh}.

\section{QCD symmetries and the $\eta$ and $\eta'$}

Symmetries are important in hadron physics.
Protons and neutrons with spin $\frac{1}{2}$ are 
related through isospin SU(2), which is expanded 
to SU(3) to include $\Sigma$ and $\Lambda$ hyperons.
Likewise, one finds SU(2) multiplets of spin-zero 
and spin-one mesons,
{\it e.g.}, 
the charged and neutral spin-zero pions are isospin 
partners and 
reside inside SU(3) multiplets together with kaons.
This spectroscopy suggests that these hadronic particles are
built from simpler constituents.
These are spin $\frac{1}{2}$ 
quarks labeled up, down and strange 
(their flavor denoted $u$, $d$ and $s$).
These quarks carry electric charges
$e_u = + \frac{2}{3}$ and $e_d, e_s = - \frac{1}{3}$
where, 
{\it e.g.}, 
a proton is built from two up quarks and a down quark, 
and a neutron is built of two down quarks and an up quark.
The spin-zero and spin-one mesons are built of a 
quark-antiquark combination.
The hadron wavefunctions are 
symmetric in flavor-spin and spatial degrees of freedom.
The Pauli principle is ensured 
with the quarks and antiquarks
being antisymmetric in a new 
label called color SU(3), red, green and blue.

High energy deep inelastic scattering experiments probe the
deep structure of hadrons by scattering high energy electron
or muon beams off hadronic targets. 
Deeply virtual photon exchange acts like a microscope which
allows us to look deep inside the proton.
One measures the inclusive cross section.
These experiments reveal a proton built of 
nearly free fermion constituents, called partons.

The deep inelastic results and spectroscopy come together
when color is made dynamical in the theory of Quantum 
Chromodynamics, QCD.
Quarks carry a color charge and interact through colored
gluon exchange, like electrons interacting through photon
exchange in Quantum Electrodynamics, QED.
QCD differs from QED in that
gluons also carry color charge whereas photons are electrically
neutral.
(The dynamics is governed by the gauge group of color
 SU(3) instead of U(1) for the photon.)
This means that the Feynman diagrams 
for QCD 
include
3 gluon and 4 gluon vertices 
(as well as the quark gluon vertices) 
and that gluons self-interact.
For excellent textbook discussions of 
QCD and its application to hadrons 
see
\textcite{Close:1979bt,Thomas:2001kw}.

Gluon-gluon interactions induce asymptotic freedom: 
the QCD version of 
the fine structure constant
for quark-gluon and gluon-gluon interactions, $\alpha_s$, 
decreases logarithmically with increasing resolution $Q^2$.
Gluon bremsstrahlung results in gluon induced jets of hadronic
particles
which were first discovered 
in high energy $e^- e^+$ collisions at DESY
\cite{Ellis:2014rma}.
Quark and gluon partons play a vital role in high energy
hadronic collisions, 
{\it e.g.}, 
at the Large Hadron Collider at CERN~\cite{Altarelli:2013tya}.
Deep inelastic scattering experiments also tell 
us that about 50\% of the proton's momentum 
perceived at high $Q^2$ is carried by gluons,
consistent with the 
QCD prediction for the deepest structure of the proton.
QCD theory also predicts that about 
50\% of the proton's angular momentum budget
is contributed by gluon spin and orbital angular momentum
\cite{Bass:2004xa,Aidala:2012mv}.

Glue in low energy QCD is manifest through the confinement
potential which binds quarks inside hadrons.
Color-singlet glueball excitations (bound states of gluons)
as well as hybrid bound states of a 
quark and antiquark 
plus gluon are predicted by theory 
but still awaiting decisive experimental confirmation.

The decay amplitude for $\pi^0 \to 2 \gamma$ and 
the ratio of cross-sections
for hadron to muon-pair production in 
high energy electron-positron collisions, $R_{e^+ e^-}$,  
are proportional to the number of dynamical colors $N_c$,
giving an experimental confirmation of $N_c =3$.

This dynamics is encoded in the QCD Lagrangian.
We first write the quark field $\psi$ 
as the sum of 
left- and right-handed quark components
$
\psi = \psi_L + \psi_R
$
where
$\psi_L = \frac{1}{2}(1 - \gamma_5) \psi$ 
and
$\psi_R = \frac{1}{2}(1 + \gamma_5) \psi$
project out different states of quark helicity.
The vector gluon field is denoted $A_{\mu}^b$.
For massless quarks, the QCD Lagrangian reads
\begin{equation}
{\cal L}_{\rm QCD}
=
{\bar \psi}_L 
i \gamma^{\mu} D_{\mu} 
\psi_L
+
{\bar \psi}_R  
i \gamma^{\mu} D_{\mu} 
\psi_R
- \frac{1}{2}
{\rm Tr} G^{\mu \nu} G_{\mu \nu} .
\ \ \ \ \ 
\\
\end{equation}
Here
$
D_{\mu} \psi = ( \partial_{\mu} - i g A_{\mu} ) \psi
$
describes the quark-gluon interaction; 
$
G^{\mu \nu} =
\partial^{\mu} A^{\nu} - \partial^{\nu} A^{\mu}
+ g f_{abc} A_b^{\mu} A_c^{\nu}
$
is the gluon field tensor with the last term 
here generating the 3-gluon and 4-gluon interactions.
The quark-gluon dynamics is determined by requiring
invariance under the gauge transformations
\begin{eqnarray}
& & \psi \to {\cal G} \psi
\nonumber \\
& & A_{\mu} \to 
{\cal G} A_{\mu} {\cal G}^{-1} 
+ {i \over g} (\partial_{\mu} {\cal G}) {\cal G}^{-1}
\end{eqnarray}
where 
${\cal G}$ 
describes rotating the local color phase of the quark fields.

For massless quarks the left- and right- handed quarks 
transform independently under chiral rotations 
which rotate 
between up, down and strange flavored quarks.
Finite quark masses 
through the Lagrangian term $m {\bar \psi} \psi$ 
explicitly breaks the chiral symmetry
by connecting left- and right-handed quarks,
\begin{equation}
{\bar \psi} \psi
=
{\bar \psi}_L \psi_R + {\bar \psi}_R \psi_L .
\end{equation}
Quark chirality 
(-1 for a left-handed quark 
 and +1 for a right-handed quark)
and helicity 
are conserved in perturbative QCD with massless quarks.

Low energy QCD is characterized by confinement and 
dynamical chiral symmetry breaking.
There is an absence of parity doublets in the 
light-hadron spectrum.
For example, 
the $J^P = \frac{1}{2}^+$ proton and 
the lowest mass $J^P = \frac{1}{2}^-$ N*(1535)
nucleon resonance 
(that one would normally take as chiral partners)
are separated in mass by 597 MeV.
This tells us that the chiral symmetry 
for light $u$ and $d$ (and $s$) quarks is spontaneously broken.

Spontaneous symmetry breaking means that the symmetry of 
the Lagrangian is broken in the vacuum. 
One finds a non-vanishing chiral
condensate connecting left- and right-handed quarks
\begin{equation}
\langle \ {\rm vac} \ | \ {\bar \psi} \psi \ | \ {\rm vac} \ \rangle < 0
.
\label{eq5}
\end{equation}
This spontaneous symmetry breaking induces an octet of 
light-mass 
pseudoscalar Goldstone bosons associated with SU(3) 
including the 
pions and kaons
which are listed in Table I
and also 
-- see below --
(before extra gluonic effects in the singlet channel)
a flavor-singlet Goldstone state.
\footnote{
Goldstone's theorem tells us that 
there is 
one massless pseudoscalar boson for
each symmetry generator that does not annihilate the vacuum.
}

The Goldstone bosons $P$ couple to the axial-vector currents 
which play the role of Noether currents 
through
\begin{equation}
\langle {\rm vac} | J_{\mu 5}^i | P(p) \rangle = 
-i f_P^i \ p_{\mu} e^{-ip.x}
\end{equation}
with $f_P^i$ the corresponding decay constants
(which determine the strength for, 
{\it e.g.}, $\pi^- \to \mu^- {\bar \nu}_{\mu}$)
and satisfy the Gell-Mann-Oakes-Renner (GMOR) relation
\cite{GellMann:1968rz}
\begin{equation}
m_P^2
f_{\pi}^2 = - m_q \langle {\bar \psi} \psi \rangle 
+ {\cal O} (m_q^2)
\end{equation}
with $f_{\pi} = \sqrt{2} F_{\pi} = 131$ MeV.
The mass squared of the Goldstone bosons $m_P^2$
is in first order proportional to the mass of their 
valence quarks, Eqs.~(5,6).
This picture is 
the starting point of successful pion and kaon phenomenology.

A scalar confinement potential implies dynamical chiral symmetry breaking. 
For example, in the Bag model of quark 
confinement is modeled by an infinite square well
scalar potential.
When quarks collide with the Bag wall, their helicity is flipped.
The
Bag wall thus connects left and right handed quarks leading 
to quark-pion coupling and the pion cloud of the nucleon 
\cite{Thomas:1982kv}.
Quark-pion coupling connected to chiral symmetry 
plays an important role in the proton's dynamics
and phenomenology,
{\it e.g.}, 
transferring net quark spin into pion cloud orbital 
angular momentum and thus playing an important role 
in the nucleon's spin structure~\cite{Bass:2009ed}.

The light mass pion is especially important in nuclear 
physics, 
also with strong coupling to the lightest mass 
$\Delta$ p-wave nucleon resonance.

\begin{table}[b!]
\begin{center}
\caption{
The octet of Goldstone bosons corresponding to chiral 
SU(3) and their masses in free space.
}
\label{bagparam}
\begin{tabular}[t]{c|ll}
\hline
Meson & wavefunction \ \ \ \ \ \ & mass (MeV) \\
\hline
$\pi^0$ & 
$\frac{1}{\sqrt{2}} (u {\bar u} -d {\bar d})$ & 135
\\
$\pi^+$     & $u \bar{d}$ & 140 
\\
$\pi^-$     & $\bar{u} d$ & 140
\\
$K^0$		& $d {\bar s}$ 			    & 498
\\
${\bar K}^0$	&	$s {\bar d}$	    & 498
\\
$K^+$  	& $u {\bar s}$  &  494
\\
$K^-$       & $\bar{u} s$  &  494
\\
$\eta_8$  
& 
$ \frac{1}{\sqrt{6}} (u\ubar + d\dbar - 2 s\sbar) $
& 
$ \frac{4}{3} m_K^2 - \frac{1}{3} m_{\pi}^2 $
\\
\hline
\end{tabular}
\end{center}
\end{table}

The QCD Hamiltonian is linear in the quark masses.
For small quark masses 
this allows one to perform a rigorous expansion
perturbing in $m_q \propto m_{\pi}^2$,
called the chiral expansion~\cite{Gasser:1982ap}.
The proton mass in the chiral limit of massless quarks
is determined by gluonic binding energy and
set by $\Lambda_{\rm QCD}$, 
which sets the scale for the running of the QCD coupling
$\alpha_s$,
$\Lambda_{\rm QCD} = 332 \pm 17$ MeV 
for QCD with 3 flavors~\cite{Patrignani:2016xqp}.

The lightest up and down quark masses
are determined from detailed studies of chiral dynamics.
One finds
$m_u=2.2^{+0.6}_{-0.4}$ MeV 
and $m_d= 4.7^{+0.5}_{-0.3}$ MeV 
whereas the strange quark mass is 
slightly heavier at $m_s= 95 \pm 5$ MeV
(with all values here quoted at the scale $\mu = 2$ GeV
 according 
 to the Particle Data Group~\cite{Patrignani:2016xqp}).

When electromagnetic interactions are also included, 
the leading order mass relations (6) become 
\cite{Georgi:1985kw}
\begin{eqnarray}
m_{\pi^\pm}^2 &=& \mu (m_u + m_d) + \Delta m^2
\nonumber \\
m_{K^\pm}^2 &=& \mu (m_u + m_s) + \Delta m^2
\nonumber \\
m_{K^0}^2 &=& \mu (m_d + m_s) 
\nonumber \\
m_{\pi^0}^2 &=& \mu (m_u + m_d) 
\nonumber \\
m_{\eta_8}^2 &=& \mu (4 m_s + m_u + m_d) 
\end{eqnarray}
where $\Delta m^2$ is the electromagnetic contribution
\cite{Dashen:1969eg} 
and
$\mu = - \langle {\bar \psi} \psi \rangle / f_{\pi}^2$.
Substituting the pion and kaon masses
gives the leading-order quark mass ratios
\begin{equation}
\frac{m_s}{m_d}\bigg|_{\rm LO} = 20, \ \ \ \ \
\frac{m_u}{m_d}\bigg|_{\rm LO} = 0.55 .
\end{equation}

The leading order GMOR formula, Eq.~(6),
gives the Gell-Mann Okubo formula 
\cite{GellMann:1961ky,Okubo:1961jc}
for the octet state
\begin{equation}
4 m_K^2 - m_{\pi}^2 = 3 m_{\eta_8}^2 .
\end{equation}
Numerically
$m_{\eta} (548 {\rm MeV}) \simeq m_{\eta_8} {\rm (570 MeV)}$.
The $\eta$ meson mass and this $\eta_8$ mass contribution 
agree within 4\% accuracy.

However, this is not the full story.
The quark condensate in Eq.~(6) 
also spontaneously breaks axial 
U(1) symmetry meaning that one might also expect a 
flavor-singlet Goldstone state which mixes with
the octet state to generate the isosinglet bosons.
However, without extra input, the resultant bosons 
do not correspond to states in the physical spectrum.
The lightest mass isosinglet bosons, the $\eta$ and 
$\eta'$, are about 300-400 MeV too heavy to be pure 
Goldstone states,
with masses $m_{\eta} = 548$ MeV and $m_{\eta'} = 958$ MeV.
One needs extra mass in the flavor-singlet channel 
to connect to the physical $\eta$ and $\eta'$ mesons.
This mass is 
associated with non-perturbative gluon dynamics.

The flavor-singlet channel is sensitive to processes
involving violation of the Okubo-Zweig-Iizuka 
(OZI) rule, 
where the 
quark-antiquark pair (with quark chirality equal two)
propagates with coupling to gluonic intermediate states 
(with zero net chirality);
see Fig.~1.
The OZI rule 
\cite{Okubo:1963fa,Zweig:1964jf,Iizuka:1966fk}
is the phenomenological observation that 
hadronic processes involving 
Feynman graphs 
mediated by gluons 
 (without continuous quark lines connecting 
 the initial and final states) 
tend to be strongly suppressed.

     \begin{figure}[t]
     \vspace{-2.5cm}
      {\centerline{  
        \includegraphics[height=0.9\textwidth]{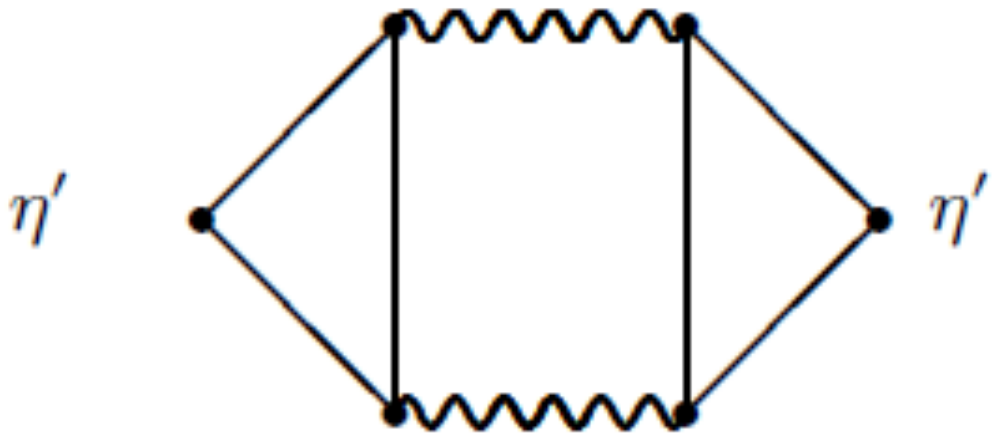}
         }}
        \vspace{-11.0cm}
       \caption{Gluonic intermediate states contribute to the $\eta'$.
The $\eta'$ mixes a chirality-two quark-antiquark 
contribution and chirality-zero gluonic contribution.
}
    \end{figure}

To see the effect of the gluonic mass contribution 
consider the $\eta$-$\eta'$ mass matrix for 
free mesons
with rows and columns in the octet-singlet basis
\begin{equation}
\eta_8 = \frac{1}{\sqrt{6}}\; (u\ubar + d\dbar - 2 s\sbar), \quad 
\eta_0 = \frac{1}{\sqrt{3}}\; (u\ubar + d\dbar + s\sbar) .
\label{mixing2}
\end{equation}
At leading order in the chiral expansion
(taking terms proportional to the quark masses $m_q$)
this reads
\begin{equation}
M^2 =
\left(\begin{array}{cc}
{4 \over 3} m_{\rm K}^2 - {1 \over 3} m_{\pi}^2  &
- {2 \over 3} \sqrt{2} (m_{\rm K}^2 - m_{\pi}^2) \\
\\
- {2 \over 3} \sqrt{2} (m_{\rm K}^2 - m_{\pi}^2) &
[ {2 \over 3} m_{\rm K}^2 + {1 \over 3} m_{\pi}^2 
+ {\tilde m}^2_{\eta_0} ] 
\end{array}\right) .
\label{eq10}
\end{equation}
Here ${\tilde m}^2_{\eta_0}$ 
is the flavor-singlet gluonic mass term.

In the notation of Eq.(7)
these singlet and mixing terms are
\begin{eqnarray}
m_{8,0}^2 &=& \mu (m_u + m_d - 2 m_s), 
\nonumber \\
m_{0}^2 &=& \mu (m_u + m_d + m_s) + {\tilde m}^2_{\eta_0} .
\end{eqnarray}

The masses of the physical $\eta$ and $\eta'$ mesons are found
by diagonalizing this matrix, {\it viz.}
\begin{eqnarray}
| \eta \rangle &=&
\cos \theta \ | \eta_8 \rangle - \sin \theta \ | \eta_0 \rangle
\nonumber \\
| \eta' \rangle &=&
\sin \theta \ | \eta_8 \rangle + \cos \theta \ | \eta_0 \rangle
\label{eq11}
\end{eqnarray}
One obtains values for the $\eta$ and $\eta'$ masses:
\begin{eqnarray}
m^2_{\eta', \eta} 
& &= (m_{\rm K}^2 + {\tilde m}_{\eta_0}^2 /2)
\nonumber \\
& & \pm {1 \over 2}
\sqrt{(2 m_{\rm K}^2 - 2 m_{\pi}^2 - {1 \over 3} {\tilde m}_{\eta_0}^2)^2
   + {8 \over 9} {\tilde m}_{\eta_0}^4} 
.
\label{eq12}
\end{eqnarray}
Here
the lightest mass state is the $\eta$ and heavier state 
is the $\eta'$.
Summing over the two eigenvalues in Eq.(14) gives the 
Witten-Veneziano mass formula
\cite{Witten:1979vv,Veneziano:1979ec}
\begin{equation}
m_{\eta}^2 + m_{\eta'}^2 = 2 m_K^2 + {\tilde m}_{\eta_0}^2 .
\end{equation}
The gluonic mass term is obtained by substituting 
the physical values of 
$m_{\eta}$, $m_{\eta'}$ and $m_K$ 
to give ${\tilde m}_{\eta_0}^2 = 0.73$GeV$^2$. 
Without the gluonic mass term
the $\eta$ would be approximately an isosinglet light-quark state
(${1 \over \sqrt{2}} | {\bar u} u + {\bar d} d \rangle$)
with mass $m_{\eta} \sim m_{\pi}$
degenerate with the pion and
the $\eta'$ would be a strange-quark state $| {\bar s} s \rangle$
with mass $m_{\eta'} \sim \sqrt{2 m_{\rm K}^2 - m_{\pi}^2}$
--- mirroring the isoscalar vector $\omega$ and $\phi$ mesons.

     \begin{figure}[t]
     \vspace{-2.5cm}
      {\centerline{  
        \includegraphics[height=0.9\textwidth]{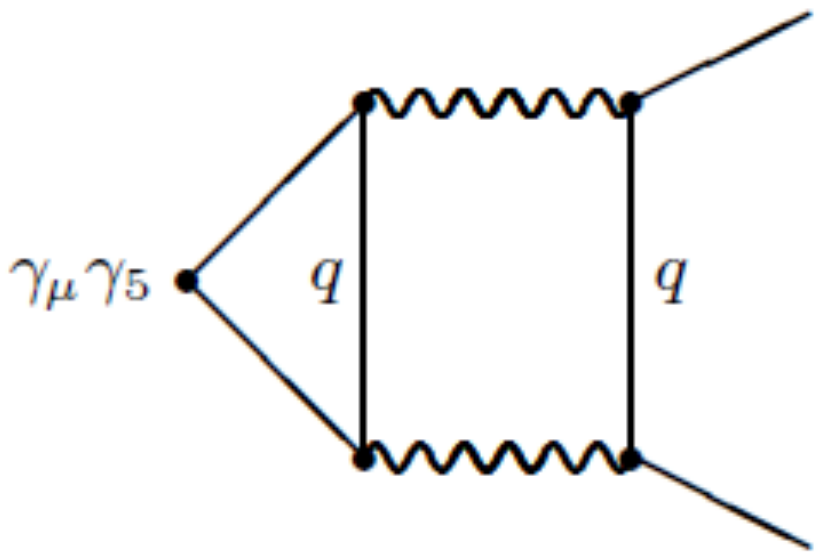}
         }}
        \vspace{-11.0cm}
       \caption{Coupling of the axial-vector current through 
gluonic intermediate states.
Gluon propagators are shown as wavy lines.  
Straight lines denote quark propagators.
}
    \end{figure}

When interpreted in terms of the leading order mixing scheme,
Eq.~(13), phenomenological studies of various decay processes 
give a value for the $\eta$-$\eta'$ mixing angle between 
$-15^\circ$ and $-20^\circ$ \cite{Gilman:1987ax,Ball:1995zv,Ambrosino:2009sc}.
The $\eta'$ has a large flavor-singlet component with 
strong affinity to couple to gluonic degrees of freedom.
Mixing means that non-perturbative glue through axial 
U(1) dynamics plays an important role in both 
 the $\eta$ and $\eta'$ and their interactions.

The gluonic mass term is associated with the QCD axial 
anomaly in the divergence of the flavor-singlet 
axial-vector current.
While the non-singlet axial-vector currents are partially conserved 
(they have just mass terms in the divergence), 
the singlet current
$
J_{\mu 5} = \bar{u}\gamma_\mu\gamma_5u
+ \bar{d}\gamma_\mu\gamma_5d + \bar{s}\gamma_\mu\gamma_5s 
$
satisfies the divergence equation 
\cite{Adler:1969gk,Bell:1969ts}
\begin{equation}
\partial^\mu J_{\mu5} = 6 Q
+ \sum_{k=1}^{3} 2 i m_k \bar{q}_k \gamma_5 q_k 
\end{equation}
where 
$
Q 
= {\alpha_s \over 8 \pi} G_{\mu \nu} {\tilde G}^{\mu \nu}
$
is called the topological charge density.
The anomalous gluonic term $Q$ is induced by QCD quantum
effects associated with renormalization of the singlet
axial-vector current
\footnote{
In QCD the flavor-singlet axial-vector
current can couple through gluon intermediate states;
see Fig.~2.
Here the triangle Feynman diagram is essential
with the axial-vector current $\gamma_{\mu} \gamma_5$
and two gluon couplings $\gamma_{\alpha}$ and $\gamma_{\beta}$
as the three vertices.
When we regularize the ultraviolet behavior of momenta 
in the triangle loop, we find that we can preserve current conservation at the quark-gluon-vertices 
(necessary for gauge invariance)
or partial conservation of the axial-vector current but 
not both simultaneously.
Current conservation wins and induces 
the gluonic anomaly term in the singlet divergence equation,
Eq.(16), from the ultraviolet point-like part of the triangle loop.
}.
Here $G_{\mu \nu}$ is the gluon field tensor and
${\tilde G}^{\mu \nu} = 
{1 \over 2} \epsilon^{\mu \nu \alpha \beta} G_{\alpha \beta}$.
For reviews of anomaly physics 
see \textcite{Shifman:1988zk} and \textcite{Ioffe:2006ww}.
Since gluons couple equally to each flavor of quark
the anomaly term cancels in the divergence equations 
for non-singlet currents like 
$J_{\mu 5}^{(3)} = 
\bar{u} \gamma_{\mu} \gamma_5 u 
- \bar{d} \gamma_{\mu} \gamma_5 d $
and 
$J_{\mu 5}^{(8)} = 
\bar{u} \gamma_{\mu} \gamma_5 u 
+ \bar{d} \gamma_{\mu} \gamma_5 d 
- 2 \bar{s} \gamma_{\mu} \gamma_5 s 
$.

The QCD anomaly means that the singlet current $J_{\mu 5}$ 
is not conserved for massless quarks.
Non-perturbative gluon processes act 
to connect left- and right-handed quarks, 
whereas 
left- and right-handed massless quarks 
propagate independently in perturbative QCD
with helicity conserved for massless quarks.

The integral over space $\int \ d^4 z \ Q = n$ 
is quantized
with either integer or fractional values
and measures a property called
the topological winding number.
This winding number vanishes in perturbative QCD and 
in QED
but is finite with non-perturbative glue,
{\it e.g.},
it is an integer for instantons 
(tunneling processes in the QCD vacuum that 
 flip quark chirality)
\cite{Crewther:1978zz}
\footnote{
For a gluon field $A_{\mu}$ with gauge transformation 
${\cal G}$,
$ 
A_{\mu} 
\to {\cal G}^{-1} A_{\mu} {\cal G} 
 + \frac{i}{g} {\cal G}^{-1} (\partial_{\mu} {\cal G}) 
$. 
Finite action requires that $A_{\mu}$ 
should tend to a pure gauge configuration 
when $x \to \infty$ with finite surface term
integral 
$\int d^4 x \ Q$ which takes 
quantized values, the topological winding number.
}.
The gluonic mass term is generated by 
glue associated with this non-trivial topology,
related perhaps to confinement or to instantons
\cite{Fritzsch:1975tx,Kogut:1974kt,Witten:1978bc,tHooft:1976rip,tHooft:1976snw}.
The exact details of this gluon dynamics are still debated.

It is interesting to consider QCD in the limit of a 
large number of colors, $N_c \to \infty$.
There are two well defined theoretical limits
taking 
$\alpha_s N_c$ and either $N_f$ (the number of flavors) 
or
$N_f/N_c$ held fixed. 
The gluonic mass term has a rigorous interpretation 
as the leading term when one makes an expansion in 
$1/N_c$
in terms of a quantity $\chi (0)$ called 
the Yang-Mills topological susceptibility,
\begin{equation}
{\tilde m}_{\eta_0}^2 \bigg|_{\rm LO}
= - {6 \over f_{\pi}^2} \chi (0) \bigg|_{\rm YM}
\end{equation}
-- for extended discussion 
see \textcite{Shore:1998dm,Shore:2007yn}.
Here
\begin{equation}
\chi (k^2)|_{\rm YM} = \int d^4 z \ i \ e^{ik.z} \
\langle {\rm vac} | \ T \ Q(z) Q(0) \ | {\rm vac} \rangle 
\big|_{\rm YM} 
\end{equation}
is calculated in the pure glue theory (without quarks).
If we assume that the topological winding number 
remains finite independent of the value of $N_c$ 
then
${\tilde m}_{\eta_0}^2 
\sim 1 / F_{\pi}^2
\sim 1 / N_c$ as $N_c \to \infty$ \cite{Witten:1979vv}.
In recent computational QCD lattice calculations
\textcite{Cichy:2015jra} 
have computed both 
the pure gluonic term on the right-hand side of Eq.(18) 
and the meson mass contributions with dynamical quarks
in the Witten-Veneziano formula Eq.(15)
and find excellent agreement at the 10\% percent level.
This calculation gives
$\chi^{1/4} (0)|_{\rm YM} = 185.3 \pm 5.6$ MeV,  
very close to the phenomenological value 180 MeV 
which follows from taking 
${\tilde m}_{\eta_0}^2 = 0.73$GeV$^2$ 
in the Witten-Veneziano formula Eq.(15).

Independent of the detailed QCD dynamics one can construct 
low-energy effective chiral Lagrangians which include the 
effect of the anomaly and axial U(1) symmetry 
\cite{DiVecchia:1980yfw,Rosenzweig:1979ay,Witten:1980sp,Nath:1979ik,Kawarabayashi:1980dp,Leutwyler:1997yr}
and use these Lagrangians to study 
low-energy processes involving the $\eta$ and $\eta'$.
We define
$
U = e^{i (  \phi / F_{\pi}
                  + \sqrt{2 \over 3} \eta_0 / F_0 ) }
$
as the unitary meson matrix 
where $\phi = \ \sum \pi_a \lambda_a$ 
denotes the octet of would-be Goldstone bosons 
$\pi_a$ associated 
with spontaneous chiral symmetry breaking
with $\lambda_a$ the Gell-Mann matrices
(SU(3) generalisations of 
 the isospin SU(2) Pauli matrices that couple to pions), 
$\eta_0$ is 
the singlet boson and 
$F_0$ is the singlet decay constant 
(at leading order 
 taken to be equal to $F_{\pi}$=92 MeV).
With this notation the kinetic energy and mass terms
in the chiral Lagrangian are
\begin{equation}
{\cal L} =
{F_{\pi}^2 \over 4}
{\rm Tr}(\partial^{\mu}U \partial_{\mu}U^{\dagger})
+
{F_{\pi}^2 \over 4} {\rm Tr} M \biggl( U + U^{\dagger} \biggr)
\end{equation}
with $M$ the meson mass matrix.
The gluonic mass term ${\tilde m}_{\eta_0}^{2}$ 
is introduced via a flavor-singlet potential involving 
the topological charge density $Q$ which is constructed 
so that the Lagrangian also reproduces the axial anomaly.
This potential reads
\begin{equation}
{1 \over 2} i Q {\rm Tr} \biggl[ \log U - \log U^{\dagger} \biggr]
+ {3 \over {\tilde m}_{\eta_0}^2 F_{0}^2} Q^2
\
\mapsto \
- {1 \over 2} {\tilde m}_{\eta_0}^2 \eta_0^2
\ \ \ \ \ 
\label{eq23}
\end{equation}
where $Q$ is eliminated through its equation of motion
to give the gluonic mass term for the $\eta'$.
The Lagrangian contains no kinetic energy term for $Q$,
meaning that the gluonic potential does not correspond
to a physical state; $Q$ is therefore distinct from mixing
with a pseudoscalar glueball state.
The $Q \eta_0$ coupling in Eq.(20)
reproduces the picture of the $\eta'$ as a mixture 
of chirality $\pm 2$ quark-antiquark and 
chirality-zero gluonic contributions; see Fig.~1.

Higher-order terms in $Q^2$ become important when we 
consider scattering processes involving more than one 
$\eta'$ or $\eta$ \cite{DiVecchia:1980vpx},
{\it e.g.}, the term 
$Q^2 \partial_{\mu} \pi_a \partial^{\mu} \pi_a$
gives an OZI-violating 
tree-level contribution 
to the decay $\eta' \rightarrow \eta \pi \pi$.
For the $\eta'$ in a nuclear medium at finite density, 
the medium dependence of ${\tilde m}_{\eta_0}^2$ 
may be introduced through coupling to the $\sigma$ 
mean field in the nucleus through the interaction term
$
{\cal L}_{\sigma Q} = g_{\sigma Q} \ Q^2 \ \sigma
$.
Here 
$g_{\sigma Q}$ denotes coupling to the $\sigma$ field.
Again eliminating $Q$ through its equation of motion,
one finds the gluonic mass term decreases in-medium
${\tilde m}_{\eta_0}^{*2} < {\tilde m}_{\eta_0}^2$ 
independent of the sign of $g_{\sigma Q}$ and the medium 
acts to partially neutralize axial U(1) symmetry breaking 
by gluonic effects \cite{Bass:2005hn}.
We return to this physics in Section VI below.
In general, couplings involving $Q$ give OZI-violation 
in physical observables.

Recent QCD lattice calculations suggest (partial) restoration
of axial U(1) symmetry at finite temperature
\cite{Bazavov:2012qja,Cossu:2013uua,Tomiya:2016jwr}.

There are several places that glue enters 
$\eta'$ and $\eta$ meson physics:
the gluon topology potential which generates the large 
$\eta'$ mass, 
possible small mixing with a lightest mass pseudoscalar 
glueball state 
(which comes with a kinetic energy term in its Lagrangian)
and, in high momentum transfer processes, radiatively 
generated glue associated with perturbative QCD.
Possible candidates for the 
pseudoscalar 
glueball state are predicted
by lattice QCD calculations with a mass above 2 GeV 
\cite{Morningstar:1999rf,Gregory:2012hu,Sun:2017ipk}.
These different gluonic contributions are distinct physics.

We have so far discussed the $\eta$ and $\eta'$ at leading 
order in the chiral expansion.
Going beyond leading order,
one becomes sensitive to extra SU(3) breaking 
through the difference in the pion and kaon decay constants,
$F_{K} = 1.22 F_{\pi}$,
as well as new OZI-violating couplings.
One finds strong mixing also in the decay constants.
Two mixing angles enter the $\eta - \eta'$ system 
when one extends the theory to $O(p^4)$ in the meson 
momentum~\cite{Leutwyler:1997yr}, 
{\it viz.}
\begin{eqnarray}
f_{\eta}^8 &=& f_8 \cos \theta_8 , \ \ \ \
f_{\eta'}^8 = f_8 \sin \theta_8 
\nonumber \\
f_{\eta}^0 &=& - f_0 \sin \theta_0 ,
\ \ 
f_{\eta'}^0 = f_0 \cos \theta_0 .
\end{eqnarray}
These mixing angles follow because the eigenstates of
the mass matrix involve linear combinations of the
different decay constants separated by SU(3) breaking
multiplying the meson states.
In the SU(3) symmetric world $F_{\pi} = F_K$ 
one would have $\theta_8 = \theta_0$,
with both vanishing for massless quarks.
One finds a systematic expansion, large $N_c$ chiral 
perturbation theory, in 
$1/N_c = O(\delta)$, $p = O(\sqrt{\delta})$ and 
$m_q = O(\delta)$, 
where $m_q$ are the light quark masses
and ${\tilde m}_{\eta_0}^2 \sim 1/N_c$.

Phenomenological fits have been made to production and 
decay processes within this two mixing angle scheme.
Best fit values quoted in \textcite{Feldmann:1999uf}
are
\begin{eqnarray}
f_8 &=& (1.26 \pm 0.04) f_{\pi}, \ \ \ \ \
\theta_8 = -21.2^\circ \pm 1.6^\circ
\nonumber \\
f_0 &=& (1.17 \pm 0.03) f_{\pi}, \ \ \ \ \
\theta_0 = -9.2^\circ \pm 1.7^\circ
\end{eqnarray}
with the fits assuming that any extra OZI-violation beyond 
${\tilde m}_{\eta_0}^2$
can be turned off in first approximation.
Similar numbers are obtained in
 \textcite{Escribano:2005qq} and \textcite{Shore:2006mm}
 and in recent QCD lattice 
 calculations~\cite{Bali:2017qce,Ottnad:2017bjt}.
To good approximation, this scheme reduces to one mixing 
angle if we change to the quark flavor basis
$\frac{1}{\sqrt{2}} (u {\bar u} + d {\bar d})$ 
and 
$s {\bar s}$, 
{\it viz.}
$\phi = 39.3^\circ \pm 1^\circ$ \cite{Feldmann:1999uf}.
These numbers 
correspond to a mixing angle about -15$^\circ$ 
in the leading order formula Eq.(13) \cite{Feldmann:1998vh}.

Recent QCD lattice calculations give values 
for the mixing angles: 
$
34 \pm 3 ^\circ$ \cite{Gregory:2011sg}
and 
$46 \pm 1 \pm 3 ^\circ$ 
\cite{Michael:2013gka,Urbach:2017rvx} in the quark flavor basis
and
$-14.1 \pm 2.8^\circ$
in the (leading order) octet-singlet basis \cite{Christ:2010dd}.

Before discussing 
phenomenology, 
we first mention two key issues 
connected to the QCD anomaly which need to be kept in mind
when understanding the $\eta'$.
Observables do not depend on renormalization scales
and are gauge invariant;
that is, they do not depend on how a theoretician 
has set up a calculation.

First,
the current $J_{\mu 5}$ picks up a dependence on the 
renormalization scale through the two-loop Feynman 
diagram in Fig.~2
\cite{Kodaira:1979pa,Crewther:1978zz}.
This means that the singlet decay constant $F_0$ 
in QCD is sensitive to renormalization scale dependence.
This is in contrast to $F_{\pi}$
which is measured by the anomaly-free current
$J_{\mu 5}^{(3)}$.
A renormalization group (RG) scale invariant version of 
$F_0$ 
suitable for phenomenology
can be defined by factoring out the scale dependence 
or, equivalently, taking the RG scale dependent quantity evaluated at $\mu^2 = \infty$.
Numerically, the RG factor is about 0.84 
if we take $\alpha_s (\mu_0^2) \sim 0.6$ 
as typical of the infrared region of QCD 
and evolve to infinity working to ${\cal O}(\alpha_s^2)$ 
in perturbative QCD \cite{Bass:2004xa}.

Second, 
the topological charge density is a total divergence
$
Q = \partial^{\mu} K_{\mu}
$.
Here
$K_{\mu}$ is the anomalous Chern-Simons current
\begin{equation}
K_{\mu} = {g^2 \over 32 \pi^2}
\epsilon_{\mu \nu \rho \sigma}
\biggl[ A^{\nu}_a \biggl( \partial^{\rho} A^{\sigma}_a
- {1 \over 3} g
f_{abc} A^{\rho}_b A^{\sigma}_c \biggr) \biggr]
\label{eqf103}
\end{equation}
with $A^{\mu}_a$ the gluon field
and $\alpha_s = g^2/4 \pi$ is the QCD coupling.
The current $K_{\mu}$ is gauge dependent.
Gauge dependence issues arise immediately if one tries 
to separate a ``$K_{\mu}$ contribution'' 
from matrix elements of the singlet current $J_{\mu 5}$.
This means that isolating the gluonic leading Fock component 
from the $\eta'$ involves subtle issues of gauge invariance 
and only makes sense with respect to a 
particular renormalization scheme like the 
gauge invariant scheme $\overline{\rm MS}$ \cite{Bass:2008fr}.

\section{The strong CP problem and axions}

The gluonic topology term (20)
which generates the gluonic 
contribution to the $\eta'$ mass 
also has the potential to 
induce strong CP violation in QCD.
One finds an extra term, $- \theta_{\rm QCD} Q$,
in the effective Lagrangian for axial U(1) physics
which ensures that the potential 
\begin{equation}
\frac{1}{2} i Q 
{\rm Tr} \biggl[ \log U - \log U^{\dagger} \biggr]
+ {3 \over {\tilde m}_{\eta_0}^2 F_{0}^2} Q^2
- \theta_{\rm QCD} Q 
\end{equation}
is invariant under axial U(1) transformations 
with
$U \rightarrow e^{-2i \alpha} U$ 
acting on the quark fields
being compensated by 
$\theta_{\rm QCD} \to \theta_{\rm QCD} - 2 \alpha N_f$.

The term $\theta_{\rm QCD} Q$ is odd under CP symmetry.
If it has non-zero value, 
$\theta_{\rm QCD}$ induces a non zero neutron
electric dipole moment \cite{Crewther:1979pi}
\begin{equation}
d_n 
= 5.2 \times 10^{-16} \theta_{\rm QCD} 
\ e {\rm cm} .
\end{equation}
Experiments constrain
$|d_n| < 3.0 \times 10^{-26}e$.cm at 90\% 
confidence limit
or
$\theta_{\rm QCD} < 10^{-10}$ \cite{Afach:2015sja}.
New and ongoing experiments
aim for an order of magnitude improvement
in precision within the next five years or so
\cite{Schmidt-Wellenburg:2016nfv}.

Why is the strong CP violation parameter $\theta_{\rm QCD}$
so small?
QCD alone offers no answer to this question.
%
%
QCD symmetries allow for a possible 
$\theta_{\rm QCD}$ 
term but do not constrain its size.
The value of $\theta_{\rm QCD}$ is an external 
parameter in the theory just like the quark masses are.

Non-perturbative QCD arguments tell us that if the 
lightest quark had zero mass, 
then there would be no net CP violation connected
to the $\theta_{\rm QCD}$ term~\cite{Weinberg:1996kr}.
However, chiral dynamics including 
the $\eta \to 3 \pi$ decay discussed 
below tells us that 
the lightest up and down 
flavor quarks have small but finite masses.
In the full Standard Model 
the parameter which determines the size of strong
CP violation is
$ \Theta_{\rm QCD} 
 = \theta_{\rm QCD} + Arg \ det \ {\cal M}_q $,
where ${\cal M}_q$ is the quark mass matrix.
Possible strong CP violation then links QCD and the 
Higgs sector in the Standard Model that determines 
the quark masses.

A possible resolution of this strong CP puzzle is to postulate
the existence of a new very-light mass pseudoscalar called the axion 
\cite{Weinberg:1977ma,Wilczek:1977pj}
which couples through the Lagrangian term
\begin{eqnarray}
{\cal L}_{a} =
&-& \frac{1}{2} \partial_{\mu} a \partial^{\mu} a
+ 
\biggl[ \frac{a}{M} - \Theta_{\rm QCD} \biggr]
\frac{\alpha_s}{8 \pi} G_{\mu \nu} {\tilde G}^{\mu \nu}
\nonumber \\
&+& 
\frac{i f_\psi}{M} \partial_{\mu} a
\ \bar{\psi} \gamma^{\mu} \gamma_5 \psi - ...
\end{eqnarray}
Here the term in $\psi$ denotes possible fermion couplings
to the axion $a$.
The mass scale $M$ plays the role of the axion decay constant 
and sets the scale for this new physics.
The axion transforms under a new global U(1) symmetry,
called Peccei-Quinn symmetry \cite{Peccei:1977hh},
to cancel
the $\Theta_{\rm QCD}$ term, 
with strong CP violation replaced by the axion coupling 
to gluons and photons.
The axion here develops a vacuum expectation value 
with the potential minimized at
${\langle {\rm vac} | a | {\rm vac} \rangle} / M 
= \Theta_{\rm QCD}$. 
The mass of the QCD axion is given by
\cite{Weinberg:1996kr}
\begin{equation}
m_a^2 = \frac{F_{\pi}^2}{M^2} 
\frac{ m_u m_d }{(m_u + m_d)^2} m_{\pi}^2 .
\end{equation}

Axions are possible dark matter candidates.
Constraints from experiments tells us that $M$ must 
be very large.
Laboratory based experiments based on the two-photon
anomalous couplings of the axion 
\cite{Ringwald:2015lqa}, 
ultracold neutron experiments to probe axion to gluon 
couplings~\cite{Abel:2017rtm},
together with astrophysics and cosmology constraints
suggest a favored QCD axion mass 
between 
$1 \mu$eV and 3 meV 
\cite{Baudis:2018bvr,Kawasaki:2013ae},
which is the sensitivity range of the 
ADMX experiment in Seattle \cite{Rosenberg:2015kxa},
corresponding to 
$M$ between about $6 \times 10^9$ and $6 \times 10^{12}$ GeV.
The small axion interaction strength, $\sim 1/M$,
means that the small axion mass corresponds to a 
long
lifetime and stable dark matter candidate,
{\it e.g.},
lifetime longer than about the present age of the Universe.
If the axions were too heavy 
they would carry too much energy out of supernova explosions, 
thereby observably shortening the neutrino arrival 
pulse length recorded on Earth in contradiction to
Sn 1987a data \cite{Kawasaki:2013ae}. 
Possible axion candidates would also need to be distinguished 
from other possible 5th force light mass scalar bosons 
\cite{Mantry:2014zsa}.

\section{$\eta$ and $\eta'$ decays}

For the $\eta$ and $\eta'$ mesons there are two main 
decay types:
hadronic decays to 3 pseudoscalar mesons and 
electromagnetic decays to two photons.
The hadronic decays are sensitive
to the details of chiral dynamics 
and, for decays into 3 pions, the
difference in the light up and down quark masses.
The two photon decays tell us about 
the spatial and quark/gluon structure of the mesons
with extra (more model dependent) information coming from
decays to $\eta'$ final states \cite{Rosner:1982ey}.
Searches for rare and forbidden decays of the $\eta$ 
and $\eta'$ mesons constrain tests of fundamental symmetries.

The total widths quoted by the Particle Data Group
are $1.31 \pm 0.05$ keV for the $\eta$ meson and 
$0.196 \pm 0.009$ MeV for the $\eta'$ \cite{Patrignani:2016xqp}
with the $\eta'$ result including 
the total width value determined directly from 
the mass distribution measured in proton-proton collisions,
$\Gamma = 0.226 \pm 0.017 \pm 0.014$ MeV
\cite{Czerwinski:2010my}.
The main branching ratios for the $\eta$ decays are
$\eta \to 3 \pi^0$ at
$32.68 \pm 0.23 \%$,
$\eta \to \pi^+ \pi^- \pi^0$ at $22.92 \pm 0.28 \%$,
and
$39.31 \pm 0.20 \%$ for the two photon decay $\eta \to 2 \gamma$.
For the $\eta'$ the main decays are
$\eta' \to \eta \pi^+ \pi^-$ at $42.6 \pm 0.7 \%$ 
and
$\eta' \to \eta \pi^0 \pi^0$ at $22.8 \pm 0.8 \%$ 
\cite{Patrignani:2016xqp}.

\subsection{Hadronic decays}

The $\eta \to 3 \pi$ decay is of key interest.
This process is driven by isospin violation in 
the QCD Lagrangian, 
the difference in light-quark up and down quark
masses $m_u \neq m_d$.
In the absence of small (few percent) 
electromagnetic contributions~\cite{Baur:1995gc}),
the decay amplitude is proportional to $m_d - m_u$
which is usually expressed in terms of the ratio
\begin{equation}
\frac{1}{R_m^2} 
= \frac{m_d^2 - m_u^2}{m_s^2 - {\hat m}^2}
\end{equation}
where ${\hat m} = \frac{1}{2}(m_d + m_u)$
and $m_s$ is the strange quark mass.
Expansion in chiral perturbation theory 
(in the light-quark masses) converges slowly due to final 
state pion rescattering effects. 
Fortunately, 
these can be resummed using dispersive techniques 
allowing 
one to make a precise determination 
of the ratio of light quark masses 
from experiments,
for a review see  
\textcite{Leutwyler:2013wna}.

Recent accurate measurements of the $\eta$ decay 
to charged pions,
$\eta \to \pi^+ \pi^- \pi^0$, 
have been performed by the WASA-at-COSY experiment 
at FZ-J\"ulich \cite{Adlarson:2014aks},
the KLOE-2 Collaboration at 
LN-Frascati \cite{Anastasi:2016cdz}
and at BES in Beijing \cite{Ablikim:2016frj}.
The neutral 3 pion decay $\eta \to 3 \pi^0$ 
has most recently been measured 
by WASA \cite{Adolph:2008vn},
KLOE \cite{Ambrosinod:2010mj},
the Mainz A2 Collaboration \cite{Prakhov:2018tou}
and at BES \cite{Ablikim:2015cmz}.

Taking the precise data on $\eta \to \pi^+ \pi^- \pi^0$ 
from KLOE-2 as input, \textcite{Colangelo:2016jmc} 
find $R_m = 22.0 \pm 0.7$. 
Combining this result with $m_s / {\hat m} = 27.30(34)$
quoted in the lattice Ref.~\cite{Aoki:2016frl},
they obtain the light quark mass ratio $m_u / m_d = 0.44 (3)$.
Similar results have been obtained 
by \textcite{Guo:2016wsi}
who include both KLOE-2 and WASA data 
for this decay and get $R_m = 21.6 \pm 1.1$.
Similar values for 
$R_m$ were found using 
earlier data by \textcite{Kampf:2011wr,Kambor:1995yc}.
These numbers compare with $R_m = 23.9$
which follows from the simple leading-order calculation
in Eq.~(8).

The decay $\eta' \rightarrow 3 \pi$ 
is also driven by isospin violation.
In addition to the QCD processes involved in the $\eta$
decay,
here there are also important contributions from the 
sub-processes
$\eta' \to \eta \pi \pi$ 
plus $\eta \pi^0$ mixing to give the 3 pion final state
and 
$\eta' \to \pi \rho$ with $\rho \to \pi \pi$.

These decays contrast with the process
$\eta' \rightarrow \eta \pi \pi$
which is the dominant $\eta'$ decay with leading QCD
term not driven by the difference in $m_u$ and $m_d$.
Here the singlet component in both the initial and final 
state isoscalar mesons $\eta'$ and $\eta$
through $\eta - \eta'$ mixing
means that the reaction is potentially sensitive
also to OZI-violating couplings, 
{\it e.g.},
from the 
$Q^2 \partial^{\mu} \pi_a \partial_{\mu} \pi_a$ 
term at next-to-leading order in $1/N_c$ in the chiral Lagrangian.
The leading order amplitude for this decay is proportional 
to $m_{\pi}^2$ and vanishes in the chiral limit.
The large branching ratios
for this decay tell us that non-leading terms play a vital role.

We refer to the lectures of \textcite{Kupsc:2009zzb}
for further details of the analysis of these processes
and 
to \textcite{Fang:2017qgz} 
for a review of the latest
experimental results from the BES experiment, 
as well as earlier measurements of these decays.

\subsection{Two-photon interactions}

The two photon decays of the $\pi^0$, $\eta$ and $\eta'$
mesons are driven by the QED axial anomaly.

For the $\pi^0$, in the chiral limit
\begin{equation}
F_{\pi} g_{\pi^0 \gamma \gamma} = {N_c \over 3 \pi} \alpha 
\end{equation}
where $g_{\pi^0 \gamma \gamma}$ 
is the $\pi^0$ two-photon coupling,
$N_c$ is the number of colors (=3)
and $\alpha$ is the electromagnetic coupling.
Without the QED anomaly the decay amplitude would be 
proportional to $m_{\pi}^2$ and vanish for massless quarks.

For the isoscalar mesons one also has to consider the 
QCD gluon axial anomaly. In the chiral limit one finds 
the relation~\cite{Shore:1991np}
\begin{equation}
F_0 \biggl[ g_{\eta' \gamma \gamma} 
    + \frac{1}{N_f} F_0 m_{\eta'}^2 g_{Q \gamma \gamma}(0) 
\biggr]
= {4 N_c \over 3 \pi} \alpha .
\end{equation}
Here $g_{\eta' \gamma \gamma}$ and $g_{Q \gamma \gamma}$
denote the two photon couplings of the physical $\eta'$ 
and topological charge density term.
Chiral corrections are discussed in \textcite{Shore:2006mm}
within the context of the two mixing angle scheme.
The observed decay rates for the $\eta$ and $\eta'$
suggest small gluonic coupling,
$g_{Q \gamma \gamma} \sim 0$, 
with the gluonic term contributing at most
10\% of the $\eta'$ decay \cite{Shore:2006mm}.
Most accurate measurements of the 
$\eta \to \gamma \gamma$
and
$\eta' \to \gamma \gamma$ decays
come from KLOE-2~\cite{Babusci:2012ik}
and BELLE~\cite{Abe:2006gn} respectively.

When one or both of the photons becomes virtual, the 
pseudoscalar meson coupling to two photon amplitudes 
involve transition form factors $F_{P \gamma} (q^2)$
associated with the spatial structure of the mesons.

There are measurements in both space-like, $Q^2 = -q^2 > 0$, 
and time-like, $q^2 > 0$, kinematics
where $q$ is the four-momentum transfer in the reaction
\footnote{Here $Q^2$ denotes the squared four-momentum
transfer of the virtual photon 
and should not be confused with ``$Q$'' 
in our previous discussion where it denoted the topological 
charge density.
For consistency with the literature we here keep $Q$ 
for both cases.}
.
The space-like region can be studied 
through $\gamma \gamma^* \to P$ fusion processes in 
electron-positron collisions, 
with $\eta$ and $\eta'$ 
production data from 
CELLO \cite{Behrend:1990sr}, 
CLEO \cite{Gronberg:1997fj},
BABAR \cite{BABAR:2011ad}
and KLOE-2~\cite{Babusci:2012ik}.
The time-like region is studied in meson decays 
$P \to \gamma \gamma^*$, $\gamma^* \to l^+ l^-$,
{\it e.g.},
Dalitz decays to lepton pairs in the final state
with positive $q^2$ equal to the invariant mass of 
the final state lepton pair $l^+ l^-$.
Single and double Dalitz decays can be studied.
Recent measurements for the $\eta$
come from the A2 Collaboration at Mainz 
\cite{Adlarson:2016hpp},
WASA-at-COSY 
\cite{Adlarson:2015zta},
and
NA60 at CERN \cite{Arnaldi:2009aa},
with data from
BES-III \cite{Ablikim:2015wnx} for the $\eta'$.

Production of a pseudocalar meson $P$
through fusion of a real and deeply virtual photon,
$\gamma \gamma^* \to P$,
are described by perturbative QCD in terms of light-front wavefunctions
\cite{Lepage:1980fj,Feldmann:1997vc}.
In the asymptotic large $Q^2$ limit, 
the transition form-factors for $\gamma \gamma^* \to P$
\begin{equation}
Q^2 F_{P \gamma} (Q^2) \to 6 \sum_a C_a f_P^a \ \ \
(Q^2 \to \infty).
\end{equation}
Here, mixing is encoded in the decay constants $f_P^a$
and 
$C_a$ are the quark charge factors.
The light-cone wavefunctions 
$\Psi_P^a (x, {\vec k}_t)$ describe the amplitude for 
finding a quark-antiquark pair carrying light-cone 
momentum fraction $x$ and $(1-x)$ and 
transverse momentum ${\vec k}_t$.
These amplitudes are normalized via
\begin{equation}
\int {d^2 {\vec k}_t \over 16 \pi^3} \int_0^1 dx 
\Psi_P^a (x, {\vec k}_t) = {f_P^a \over 2 \sqrt{6}} .
\end{equation}
As we explained in Section II, one cannot separate an anomalous $K_{\mu}$ contribution from $F_0$ 
when working with gauge invariant observables, 
{\it e.g.}, using $\overline{\rm MS}$ renormalization.
The small OZI-violation in $F_0$ is consistent 
with RG effects and with the quark-antiquark leading 
Fock component moving in a topological gluon potential.
Glue may be (strongly) excited in the intermediate states of
hadronic reactions.

The low $q^2$ region is described using form-factors
\begin{equation}
F (q^2) = F(0)
\frac{\Lambda^2}{\Lambda^2 - q^2 - i \Gamma \Lambda} .
\end{equation}
The slope parameter 
\begin{equation}
b_P = \frac{d |F (q^2)|}{dq^2} \Big|_{q^2=0} = 
F(0) \frac{1}{\Lambda^2 + \Gamma^2} 
\end{equation}
is often quoted for the decays.
Values extracted for the $\eta'$ from timelike decays 
are $b_{\eta'} = 1.60 \pm 0.17 \pm 0.08$ GeV$^{-2}$ 
and $\Lambda = 0.79 \pm 0.04 \pm 0.02$ GeV
from BES-III~\cite{Ablikim:2015wnx},
with $\Lambda$ close to the $\omega$ and $\rho$ masses 
which appear with vector meson dominance of the virtual photon.
In the space-like region the CELLO Collaboration found
$b_{\eta'} = 1.60 \pm 0.16$ GeV$^{-2}$ \cite{Behrend:1990sr}.
Note that the $\Gamma$ width term 
is important here for the $\eta'$ 
because of its large mass and short life time.
For the $\eta$ slope measured in time-like decays, 
the most precise measurement of $\Lambda_{\eta}^{-2}$ is
$1.97 \pm 0.11$ GeV$^{-2}$ 
from the A2 Collaboration at Mainz~\cite{Adlarson:2016hpp}.

\begin{figure}[h]
     \vspace{-3.0cm}
      {\centerline{  
        \includegraphics[height=0.27\textwidth]{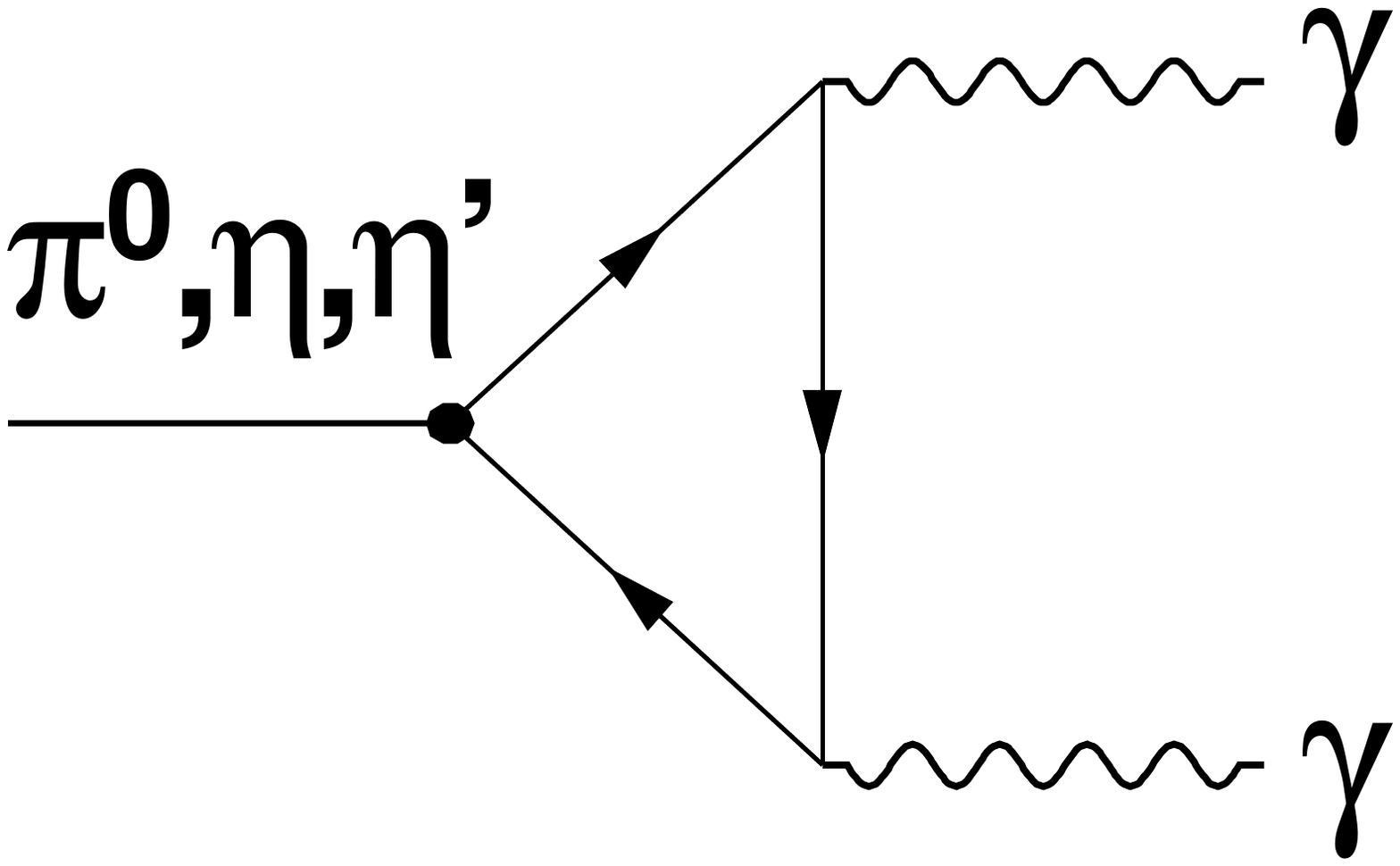}
        \includegraphics[height=0.27\textwidth]{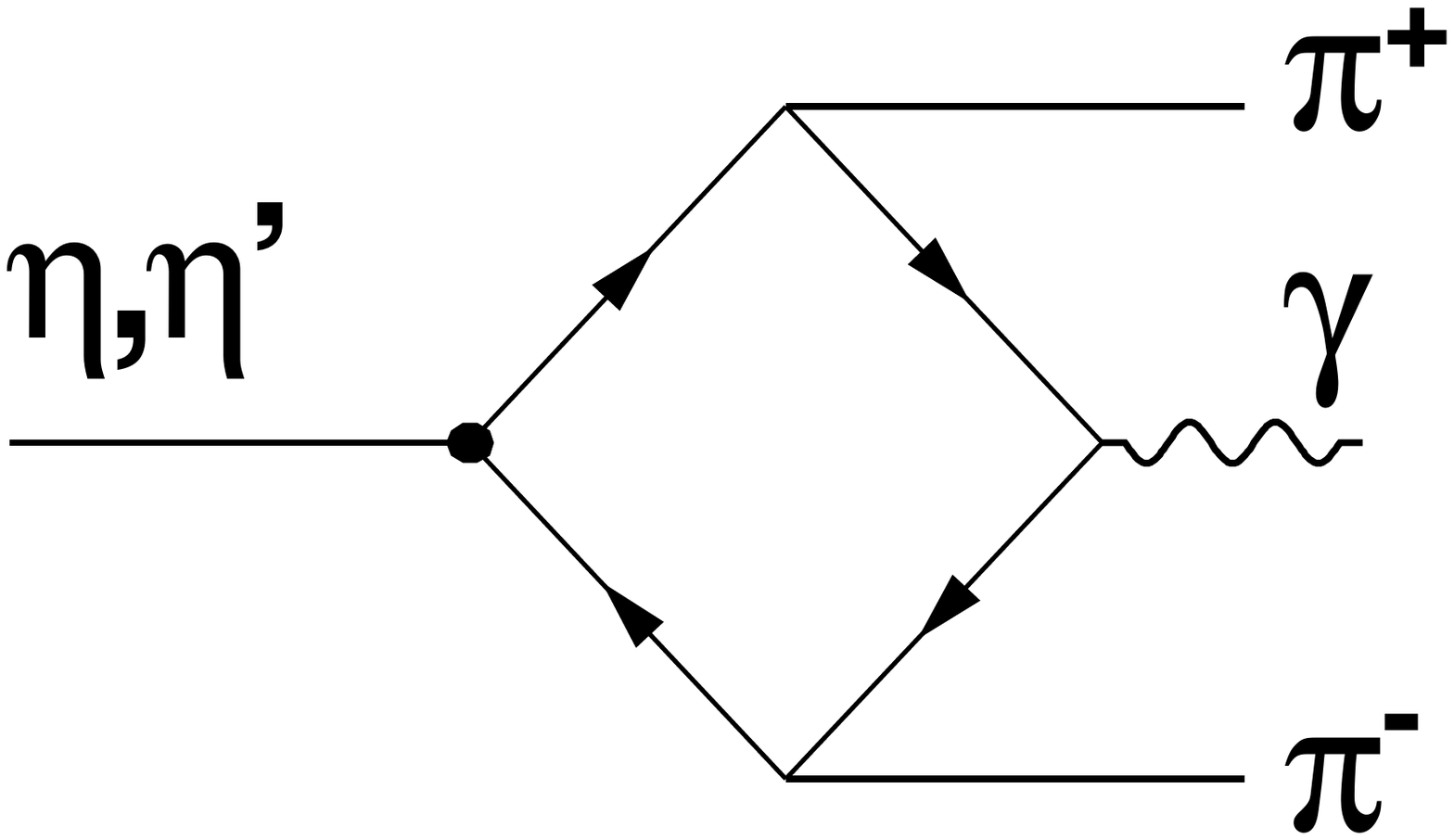} }}
       \caption{Feynman diagrams for the triangle and box anomalies. The anomaly comes from the point-like part of
the quark loop with the quarks carrying maximum momentum
in the loop.
               \label{box_triangle}}
    \end{figure}

Extending the final states from charged leptons 
to charged pions, the process $\eta \to \pi^+ \pi^- \gamma$
includes contributions from both the transition 
form-factor and also the box anomaly 
shown in Fig.~\ref{box_triangle}.
Recent measurements 
are from WASA \cite{Adlarson:2011xb}
and KLOE-2 \cite{Babusci:2012ft}.
For a recent theoretical discussion see \textcite{Kubis:2015sga}.

The $\eta' \gamma$ transition form factor 
for deeply virtual
$\gamma^* \gamma \rightarrow  \eta'$ 
was interpreted in \cite{Kroll:2013iwa} 
to give a quite large (radiatively generated)
two gluon Fock component in the $\eta'$ wavefunction. 
In this calculation the glue enters at next-to-leading order. 
Exclusive central production of the $\eta'$ in 
high-energy proton-proton collisions
at the LHC has been suggested as a cleaner 
probe since here the glue enters at leading-order 
\cite{Harland-Lang:2013ncy}.

In lower energy experiments, quark model inspired 
fits including a ``gluonium admixture''
\cite{Rosner:1982ey}
have been performed to various low energy processes 
including the $\phi \rightarrow \eta' \gamma$ decay
by the KLOE Collaboration 
\cite{Ambrosino:2009sc,Ambrosino:2006gk,Gauzzi:2012zz}
suggesting a phenomenological ``gluonium fraction'' 
of $0.12 \pm 0.04$.
Various theoretical groups' analyses of the same data
suggest values between zero and about 10\% depending
on form-factors that are used in the fits 
\cite{Thomas:2007uy,Escribano:2007cd,DiDonato:2011kr}.
When trying to extract a ``gluonic content'' 
from experiments it is important to be careful 
what assumptions about glue have gone into the analyses.
Photon coupling decay processes are theoretically 
cleaner with less model dependence in their interpretation.

At high energies, 
heavy-quark meson decays to light-quark states including 
the $\eta'$ proceed through OZI-violating gluonic 
intermediate states,
{\it e.g.}, 
$J/\Psi$ to $\eta' \gamma$ and $\eta \gamma$
giving experimental constraints on the flavor-singlet
components in these mesons.
In high energy processes large branching ratios 
for $D_s$ and $B$ meson decays 
to $\eta'$ final states have been observed 
and are believed to be driven in part by coupling to gluonic intermediate states 
\cite{Browder:1998yb,Aubert:2001sr,Behrens:1998dn,Ball:1995zv,Fritzsch:1997ps,Atwood:1997bn,Hou:1997wy,Bali:2014pva,
Dighe:1995gq,Dighe:1997hm}.

\subsection{Precision tests of fundamental symmetries}

Precision measurements of the muon's anomalous magnetic 
moment $a_{\mu} = (g-2)/2$ 
are an important test of the Standard Model.
The anomalous magnetic moment is induced 
by quantum radiative corrections to the
magnetic moment 
with $g$ the proportionality constant 
between the particle's magnetic moment and its spin. 
The present experimental value from BNL~\cite{Bennett:2006fi}
\begin{equation}
a_{\mu}^{\rm exp} 
= (11659209.1 \pm 5.4 \pm 3.3) \times 10^{-10}
\end{equation}
differs from the present best theoretical expectation by
\begin{equation}
a_{\mu}^{\rm exp} - a_{\mu}^{\rm th}
=
(31.3 \pm 7.7) \times 10^{-10}
\end{equation}
-- a 4.1 $\sigma$ deviation \cite{Jegerlehner:2017gek}.
This result is a puzzle also since possible new physics
contributions which might have resolved the discrepancy 
are now seriously challenged by LHC data which are, 
so far, consistent with the Standard Model and 
no extra new particles in the mass range of the experiments.
New experiments at Fermilab and J-PARC plan to check this 
result with the Fermilab experiment
improving the present statistical error on $a_{\mu}$ 
from 
540 to 140 ppb or $1.4 \times 10^{-10}$~\cite{Hertzog:2015jru}.

One key issue is the size of low-energy QCD hadronic
contributions to the muon $g-2$. 
These are the biggest source of theoretical 
uncertainty in the Standard Model prediction
with one important ingredient being 
the hadronic contributions 
to virtual photon-photon scattering with meson intermediate states.
These are sensitive to the $\pi^0$, $\eta$ and $\eta'$ 
transition form-factors.
Various calculations appear in the literature;
see Table 5.13, page 474, 
in \textcite{Jegerlehner:2017gek}.
Contributions to $a_{\mu}$ from the $\eta$ and $\eta'$ 
are typically about $3 \times 10^{-10}$ with 
pion contributions between about 5 and 8 $\times 10^{-10}$.
The total hadronic contribution to $a_{\mu}$ 
including vacuum polarization effects is about 
690 $\times 10^{-10}$ 
with a net light-by-light contribution of about
10 $\times 10^{-10}$ 
after summing over terms with positive and negative signs.

Studies of $\eta$ meson decays also provide new precision 
tests of discrete symmetries: 
charge conjugation, $C$, 
and charge-parity, $CP$ \cite{Jarlskog:2002zz}.
The $\eta$ and $\eta^{\prime}$ mesons are eigenstates of 
parity $P$, charge conjugation and combined $CP$ parity 
with eigenvalues $P$~=~-1, $C$~=~+1 and $CP$~=~-1.
$C$ tests include searches for forbidden decays to an odd 
number of photons.
{\it e.g.}, 
$\eta \to 3\gamma$ \cite{Nefkens:2005ka},
$\eta \to \pi^0\gamma$ 
(which is also forbidden by angular momentum conservation)
\cite{Adlarson:2018imw}, 
and $\eta \to 2\pi^0 \gamma$ \cite{Nefkens:2005dp}.
Charge conjugation  invariance has also been tested in 
the $\eta \to \pi^0 \pi^+ \pi^-$ decay.
Here $C$ violation can manifest itself as an asymmetry in 
the energy distributions for $\pi^+$ and $\pi^-$ mesons 
in the rest frame of the $\eta$ meson. 
The results were found consistent with zero 
\cite{Adlarson:2014aks}.
A possible $CP$ violating asymmetry in the
$\eta \to \pi^+ \pi^- e^+ e^-$ decay was 
determined consistent with zero \cite{Adlarson:2015zta}.

\section{$\eta$- and $\eta'$-nucleon interactions}

Close-to-threshold $\eta$ and $\eta'$ production is 
studied in photon-nucleon and proton-nucleon collisions.
Photon induced reactions are important for studies of 
nucleon resonance excitations;
for a recent review see \textcite{Krusche:2014ava}.
$\eta$ meson production is characterized by the strong 
role of the s-wave N*(1535) resonance.
For studies of higher mass excited resonances, recent 
advances with double polarization observables are playing
a vital role.
Recent measurements for the $\eta$
come from Mainz
\cite{Witthauer:2017wdb,Witthauer:2017get}, 
Jefferson Laboratory \cite{AlGhoul:2017nbp,Senderovich:2015lek} 
and GrAAL \cite{Sandri:2014nqz},
with partial wave analysis 
studies reported in \textcite{Anisovich:2015tla}.

For the $\eta'$, (quasi-free) photoproduction from proton and 
deuteron targets has been studied at 
ELSA \cite{Crede:2009zzb,Jaegle:2010jg,Krusche:2012zza},
MAMI \cite{Kashevarov:2017kqb}
and by the CLAS experiment at Jefferson Laboratory 
\cite{Dugger:2005my,Williams:2009yj}
with new double polarization observables reported 
in \textcite{Collins:2017sgu}.
The production cross-section is isospin independent
for incident photon energies greater than 2 GeV, 
where $t$-channel exchanges are important.
At lower energies, particularly between 1.6 and 1.9 GeV
where the proton cross-section peaks, the proton and
quasi-free neutron cross-sections show different
behavior.
These data have recently been used in partial wave analysis 
revealing strong indications of 
four excited nucleon resonances contributing to the $\eta'$
production process:
$N(1895) \frac{1}{2}^-$,
$N(1900) \frac{3}{2}^+$,
$N(2100) \frac{1}{2}^+$, and 
$N(2120) \frac{3}{2}^-$. 
Details including the branching ratios for coupling 
to the $\eta'$ are given in \textcite{Anisovich:2017pox}.

In proton-nucleon collisions
the $\eta$ and $\eta'$ production processes proceed through exchange of 
a complete set of virtual meson hadronic states, 
which in models is usually truncated to single 
virtual meson-exchange,
{\it e.g.}, 
$\pi$, $\eta$, $\rho$, $\omega$ and $\sigma$ 
(correlated two-pion) exchanges
\cite{Faldt:2001uz,Nakayama:2003jn,Pena:2000gb,
Deloff:2003te,Shyam:2007iz}.
For the $\eta'$ OZI-violating production 
is also possible through excitation of 
non-perturbative glue in the interaction 
region~\cite{Bass:1999is}.
The exchange process can also induce nucleon resonance
excitation, especially the N*(1535) with $\eta$ production,
before final emission of the $\eta$ or $\eta'$ meson.
The production mechanism 
is studied through measurements of the total and differential
cross-sections, 
varying the isospin of the second nucleon and 
polarization observables with one of the incident protons
transversely polarized \cite{Moskal:2004cm}.
The interpretation of these processes is sensitive 
to the choice of exchanged mesons and nucleon resonances
included in the models and 
the truncation of the virtual exchange contributions 
which affects, 
{\it e.g.}, the meson nucleon form-factors in the calculations.

The near-threshold $\eta$ meson production in nucleon-nucleon
collisions has been investigated extensively in the 
CELSIUS, COSY and SATURNE facilities.
The results determined by different experiments
for the 
total~\cite{Bergdolt:1993xc,Hibou:1998de,Chiavassa:1994ru,Calen:1996mn,Smyrski:1999jc,Moskal:2003gt,Moskal:2009bd}  
and differential~\cite{Moskal:2003gt,AbdelBary:2002sx,Moskal:2009bd,Petren:2010zz}
cross-sections for the $pp \to pp\eta$  
and for the quasi-free $pn \to pn\eta$ reactions~\cite{Calen:1996mn,Calen:1998vh,Moskal:2008pi}
are consistent within the estimated uncertainties.
In the different experiments $\eta$ mesons could be
produced up to excess energy ${\cal E}$
of 92 MeV at CELSIUS, 502 MeV at COSY and 593 MeV at SATURNE.

$\eta'$ production has been measured in proton-proton
collisions close-to-threshold 
(excess energy ${\cal E}$ between 0.76 and $\sim 50$ MeV) 
by the COSY-11 collaboration at FZ-J\"ulich 
\cite{Moskal:1998pc,Moskal:2000gj,Moskal:2000pu,Khoukaz:2004si,
Czerwinski:2014ama,Klaja:2010vy}
and at ${\cal E} = 3.7$ MeV and 8.3 MeV 
by SPESIII \cite{Hibou:1998de} 
and 
144 MeV by the DISTO Collaboration at SATURNE 
\cite{Balestra:2000ic}.

For near-threshold meson production,
the cross-section is reduced by initial state interaction
between the incident nucleons and enhanced by final state
interactions between the outgoing hadrons. 
For comparing production dynamics a natural variable is
the volume of available phase space 
which is approximately independent of the meson mass.
Making this comparison for the neutral pseudoscalar mesons, 
it was found that production of the $\eta$ meson is about 
six times enhanced compared to the $\pi^0$ which is six 
times further enhanced compared to the $\eta'$. 
The production amplitudes for the $\pi^0$ and $\eta'$
have the same (nearly constant) dependence on the phase
space volume in the measured kinematics close-to-threshold,
whereas the production amplitude for the $\eta$ 
exhibits possible 
growth with decreasing phase space volume 
due to strong $\eta$-proton attractive 
interaction~\cite{Moskal:2000pu}.
The large $\eta$ production cross-section is driven by strong coupling to the N*(1535).
\begin{figure}[t!]  
\vspace{-1cm}
\centerline{\includegraphics[width=0.37\textwidth]
{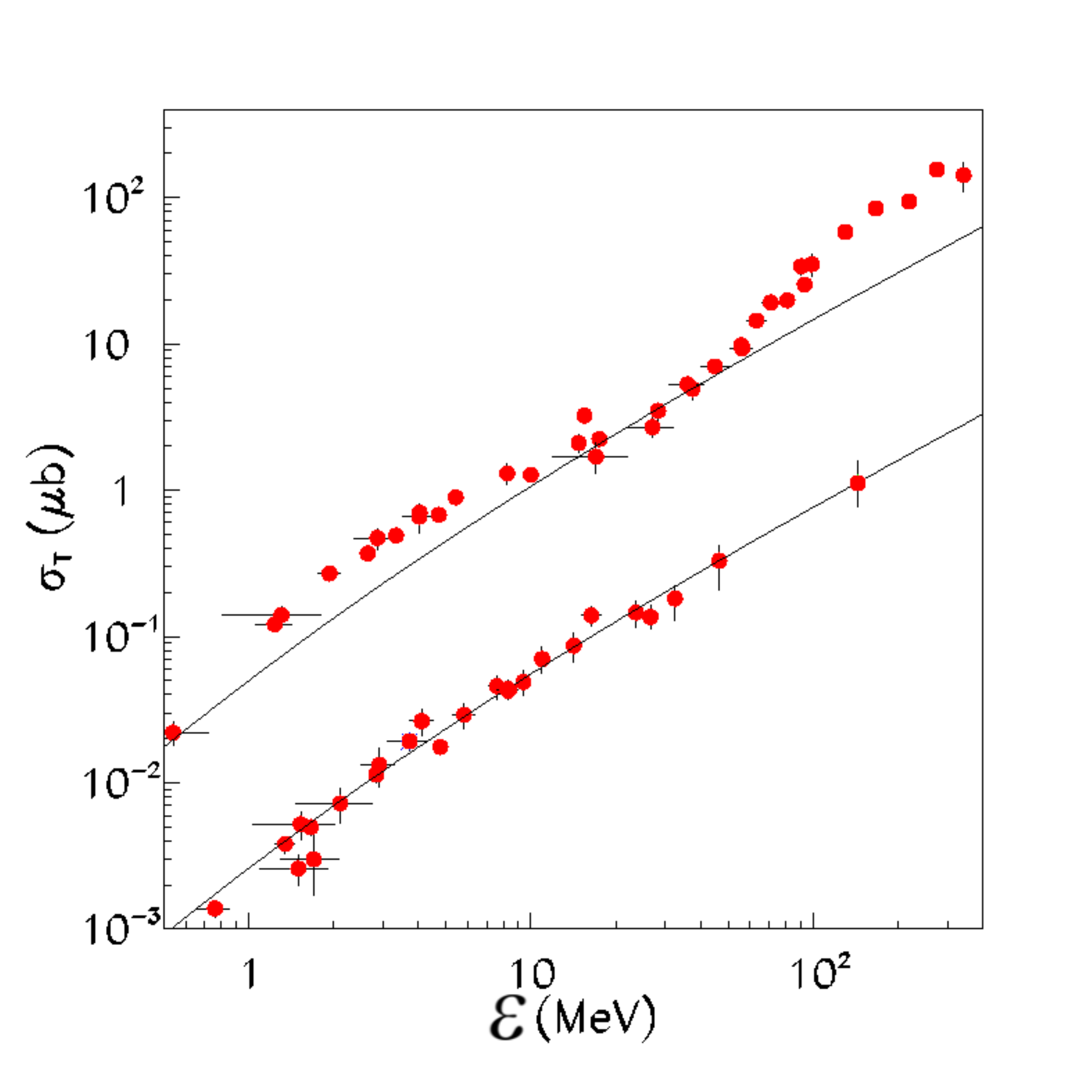}}
\caption{ 
\label{fig:eta_etaprime} 
World data for the total cross sections for $pp\to pp\eta$ 
(upper points) and $pp\to pp\eta^{\prime}$ (lower points)
-- see text. 
The solid curves are arbitrarily scaled $pp$ FSI
predictions of Eq.~(\ref{FW1}).
The Figure is adapted from~\textcite{Wilkin:2016ajz}. 
}
\end{figure}
In Fig.~\ref{fig:eta_etaprime} 
we show the $\eta$ and $\eta'$ production total
cross-section data as a function of excess energy.
The Figure also shows the curves expected if 
one includes only the $s-$wave and 
final state interaction in the proton-proton in the simplest 
approximation~\cite{Wilkin:2016ajz,Faeldt:1996na}:
\begin{equation}
\label{FW1}
\sigma_T(pp\to pp\eta)
=C \left.\left(\frac{{\cal E}}{\mu}\right)^2\right/
   \left(1+\sqrt{1+{\cal E}/\mu}\right)^{2} .
\end{equation}
Here the excess energy 
${\cal E} =W -(2m_p + m)$, 
with $W$ the total center of mass energy,
$m_p$ the proton mass and $m$ the meson mass. 
The constant $C$ depends upon the reaction mechanism and 
can be adjusted to fit the data.
Strong $\eta$-nucleon final state interaction is seen at
the lowest ${\cal E}$ with deviation of the data from 
the theoretical curve,
much stronger than for the $\eta'$.
Deviations at large ${\cal E}$ are likely to originate 
from higher partial waves in the final proton-proton system.
The pole parameter $\mu$ fitted from experiment is 
$\approx 0.75$ MeV for the $\eta'$~\cite{Wilkin:2016ajz}.

Measurements of the differential cross-sections for $\eta$
production at 
${\cal E} = 15.5$ MeV \cite{Moskal:2003gt}
and at ${\cal E} = 41$ MeV \cite{AbdelBary:2002sx}
are consistent with isotropic 
$\eta$ production within the statistical errors,
though at 41 MeV the accuracy of the data do not 
exclude a few per cent contribution from higher partial waves.
For $\eta'$ production, the differential cross sections 
measured at 
SATURNE~\cite{Balestra:2000ic} at ${\cal E} = 143.8$ MeV 
and at 
COSY~\cite{Khoukaz:2004si} at ${\cal E} = 46.6$ MeV 
are consistent with pure $Ss$-wave production
with $\approx 10 \%$ level higher partial wave 
contributions possible within the experimental uncertainties.
Here $Ss$ denotes the outgoing protons in $S$-wave 
in their rest frame and the meson is in $s$-wave relative
to the center-of-mass.

Values for the real part of the $\eta$-nucleon
scattering length $a_{\eta N}$ have been obtained 
between 0.2 fm and 1.05 fm depending on the analysis, 
including whether the $\eta - \eta'$ mixing angle is 
constrained or not.
Fits to experimental data suggest a value close 
to 0.9 fm for the real part of $a_{\eta N}$ 
\cite{Green:1999iq,Green:2004tj,Arndt:2005dg}.
In contrast, smaller values of $a_{\eta N}$ with real part 
$\sim 0.2$ fm are predicted by chiral coupled-channel
models where 
the $\eta$ meson is treated in pure octet approximation 
\cite{Waas:1997pe,Inoue:2002xw,GarciaRecio:2002cu}.

The scattering length $a_{\eta N}$ is much greater than the
scattering length for pion-nucleon scattering.
Pion nucleon interactions are 
dominated by the $p$-wave $\Delta$ (lightest mass) 
nucleon
resonance excitation with small scattering length,
which for the $\pi^0$ the real part is
$a_{\pi N} = 0.1294 \pm 0.0009$~fm~\cite{Sigg:1996qd}.

The COSY-11 collaboration have recently made a first
measurement of the $\eta'$-nucleon scattering length 
in free space,
\begin{equation}
a_{\eta' p} = 
(0 \pm 0.43) + i (0.37^{~+0.40}_{~-0.16})~\mathrm{fm}
\label{Eq:C11-a-eta-prime}
\end{equation}
from studies of the $\eta'$ final state interaction in 
$\eta'$ production in proton-proton collisions 
close-to-threshold~\cite{Czerwinski:2014yot}. 
This value was extracted from fitting the low ${\cal E}$
data, ${\cal E}$ up to 11 MeV, 
where the cross-section is clearly s-wave dominated.
A recent extraction from photoproduction data gives
\begin{equation}
|a_{\eta' N}| = 0.403 \pm 0.015 \pm 0.060 ~\mathrm{fm}
\end{equation}
with phase $87 \pm 2^\circ$ 
\cite{Anisovich:2018yoo}.
Theoretical models in general prefer a positive sign for 
the real part of $a_{\eta' p}$ corresponding to attractive interaction.
The meson-nucleon scattering lengths are also related 
to the corresponding meson-nucleus optical potential;
see Section VI below. 
Measurements of the $\eta'$ mass shift in carbon 
favor a value for the real part of $a_{\eta' N}$ of 
about 0.5~fm.

These numbers can be understood in terms of the underlying dynamics.
In chiral dynamics, the Goldstone-boson nucleon scattering lengths are proportional at tree level to the 
meson mass squared, 
{\it e.g.}, the Tomozawa-Weinberg relation
\cite{Ericson:1988gk}.
For pion-nucleon scattering, 
the nearest $s$-wave resonance is the N*(1535), 
which is too far away to affect the near-threshold interaction.
For the $\eta$ one finds a strong effect from the 
close-to-threshold resonance N*(1535).
With the $\eta'$, the meson mass squared is large through
the gluonic mass term ${\tilde m}^2_{\eta_0}$.
The tree level scattering length is non-vanishing in the
chiral limit.

Measurements of the isospin dependence of $\eta$ meson 
production in proton-nucleon collisions 
revealed that the total cross-section 
for the quasi-free $pn \rightarrow pn \eta$ reaction
exceeds the corresponding cross section 
for $pp \rightarrow pp \eta$ 
by a factor of about three at threshold and by factor of 
six at higher excess energies between about 25 and 100 MeV
\cite{Moskal:2008pi,Calen:1998vh}.
The strong isospin dependence tells us there must be a 
significant isovector exchange contribution at work in
the proton-nucleon collisions.

The spin analyzing power $A_y$ for $\eta$ meson production 
in proton-proton collisions close-to-threshold with one 
proton beam transversely polarized 
has recently been measured with high statistics
by the WASA-at-COSY Collaboration~\cite{Adlarson:2017jtw}.
The analyzing power is found to be consistent with zero 
for an excess energy of ${\cal E} = 15$ MeV 
signaling $s$ wave production with no evidence for
higher partial waves. 
This result is in contrast with meson-exchange model 
predictions 
which had anticipated asymmetries up to about 
20 \% 
based on $\pi$ or $\rho$ exchange dominance in
the interaction
\cite{Faldt:2001uz,Nakayama:2003jn};
see Fig.~5.
At ${\cal E} = 72$ MeV the data reveal strong interference 
of $Ps$ and $Pp$ partial waves and cancellation of $(Pp)^2$ and 
$Ss*Sd$ contributions~\cite{Adlarson:2017jtw}.
Different meson-exchanges induce very different spin 
dependence in the production process.
Polarized beams and measurement of the analyzing power 
can therefore put powerful new constraints on theoretical
understanding of the $\eta$ production process.
A possible explanation of the vanishing analyzing power at
15 MeV might be cancellation with destructive interference 
between $\pi$ and $\rho$ exchanges in $\eta$ production 
very close-to-threshold together with a 
strong (spin independent) scalar $\sigma$ 
(correlated two pion) exchange contribution.
In this scenario one would expect to see a finite analyzing
power in proton-neutron collisions given the strong isospin
dependence to the production mechanism.

\begin{figure}[t!]   
\vspace{-8cm}
\mbox{\hspace{-0.2cm}\includegraphics[width=0.54\textwidth]{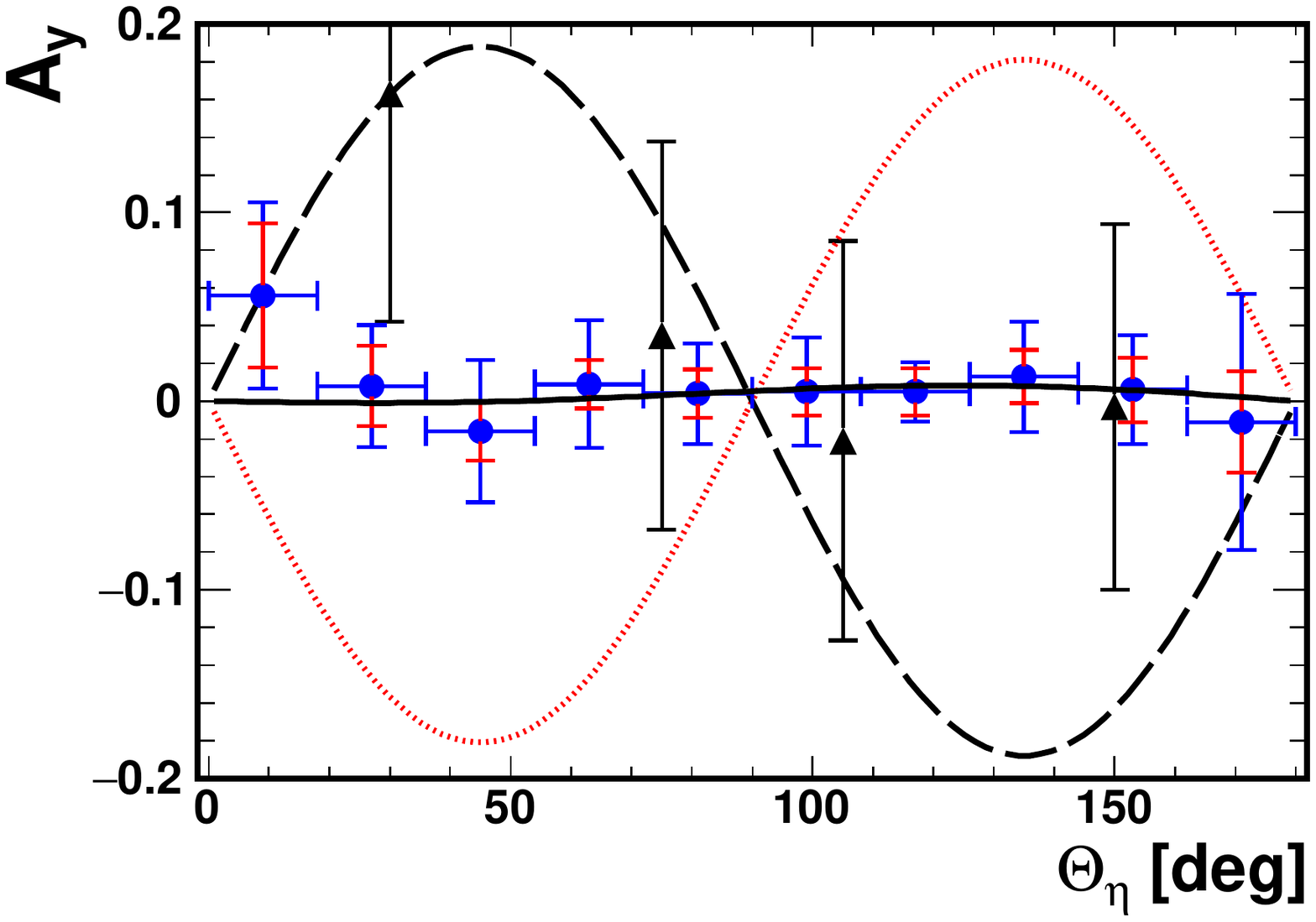}}
\caption{
Analyzing power for the $\vec{p}p\to pp\eta$ reaction at 
Q~=~15~MeV.
Here $\theta_{\eta}$ is the polar angle for the 
emission of the $\eta$ meson in the 
center of mass system. 
Full circles represent 
WASA results~\cite{Adlarson:2017jtw}.
Triangles are early data from COSY-11 
measured at
${\cal E}$~=10~MeV~\cite{Czyzykiewicz:2006jb}.
The dotted line denotes the prediction based
on pseudoscalar-meson-exchange \cite{Nakayama:2003jn},
whereas the dashed line represents the vector exchange model
 \cite{Faldt:2001uz}.
The solid line is the partial-waves fit to the WASA data.
The Figure is adapted from \textcite{Adlarson:2017jtw}.
\label{fig:AyQ15}
}
\end{figure}

Measurements of the isospin dependence of $\eta'$ 
production 
suggest a different production
mechanism for this meson \cite{Moskal:2000pu,Klaja:2010wk}.
Using the quasi-free proton-neutron interaction 
\cite{Moskal:2005uh}
COSY-11 placed an upper bound on 
$\sigma (pn \rightarrow pn \eta')$ and the ratio
$R_{\eta'} = 
\sigma (pn \rightarrow pn \eta') 
/ \sigma(pp \rightarrow pp \eta')$ \cite{Klaja:2010wk}.
For excess energy between 8-24 MeV 
the upper limit of 
$R_{\eta'}$
was observed to be consistently one standard 
deviation below the corresponding ratio for $\eta$
production \cite{Moskal:2008pi}. 
In the theoretical limit that $\eta'$ production proceeds
entirely through gluonic excitation in the intermediate
state this ratio would go to one. 
The data are consistent with both a role for OZI-violating 
$\eta'$ production \cite{Bass:1999is,Bass:2000ie} 
and the meson-exchange model \cite{Kaptari:2008ae}.

The observed $s$-wave dominance of $\eta$ and $\eta'$ 
production in a large kinematic range 
close-to-threshold might also, in part, be understood 
in terms of the phenomenology of \textcite{GellMann:1954wr}.
If the strength of the primary production partial 
amplitudes were constant over the phase space,
then the energy dependence of the partial cross sections
would be given by
\begin{equation}
\label{sigmaLl}
\sigma_{Ll} 
\propto \mbox{q}_{\,max}^{\,2L+2l+4} 
\propto \eta_{\,M}^{\,2L+2l+4} .
\end{equation}
Here $\eta_{M} = \mbox{q}_{max}/m$ 
with $m$ and $\mbox{q}_{\,max}$ 
the mass and maximum momentum of the created meson.
Close-to-threshold the $Ss$ partial-wave cross-section 
should increase with the fourth power of $\eta_{M}$
which, non-relativistically, 
is related to the excess energy by
${\cal E} = \eta_M^2 m (2m_p + m) / 4 m_p$.
The orbital angular momentum $l$ of the produced meson 
is $l = R\,q \sim \,q/ m$,
where $R$ is a characteristic distance from the 
center of the collision 
$R \sim 1 / m  \sim 1 /\Delta p$ 
with $m$ the meson mass and $\Delta p$ 
the momentum transfer between the colliding nucleons. 
Hence $\eta_M$ denotes the classically calculated 
maximum angular momentum of the meson in the center of mass frame.

Investigations with polarized beams and 
targets~\cite{Meyer:2001gj,Meyer:1999vf}
of the $\vec{p}\vec{p} \to pp\pi^{0}$ reaction tell us 
that the $Ss$ partial-wave accounts for more than $95\,\%$ 
of the total cross section up to $\eta_M \approx 0.4$. 
Extending this phenomenology to heavy mesons suggests that
the $Ss$ partial wave combination will constitute the 
overwhelming fraction of the total production 
cross-section for $\eta_M$ smaller than about 0.4
for constant production amplitudes $|M_{Ll}^0|$.
That is, one expects the heavier $\eta$ and $\eta'$
mesons to be produced predominantly via the $Ss$ state 
in a much larger excess energy range and hence larger 
phase space volume. 
Whereas for $\pi^0$ production the onset of higher partial 
waves is observed at 
${\cal E}$ around $10\,\mbox{MeV}$, 
it is expected only above $100\,\mbox{MeV}$ for the $\eta'$
and
above  $\approx 40\,\mbox{MeV}$ for the $\eta$ meson
(modulo the possible small 
 change in amplitude with increasing 
phase space volume~\cite{Klaja:2010vy,Moskal:2000pu}).

\subsection{The N*(1535) resonance and its structure}

The internal structure of the N*(1535) has been a hot topic
of discussion.
In quark models the N*(1535)
is interpreted as a 3-quark state: $(1s)^2(1p)$.
One finds configuration mixing with the N*(1650)
between 
$|^2 P_{\frac{1}{2}} \rangle$ and 
$|^4 P_{\frac{1}{2}} \rangle$ states 
(with spin $\frac{1}{2}$ and $\frac{3}{2}$ 
 respectively, orbital angular momentum $L=1$ and
 total angular momentum $J=\frac{1}{2}$) 
\cite{Isgur:1978xj}. 
Recent QCD lattice calculations support a 
3-quark state, with couplings to 5 quark components
and probability of about 50\% to contain the bare baryon
\cite{Liu:2015ktc}.
This contrasts with the $\Lambda (1405)$ resonance which is
understood as 
dynamically generated in the kaon-nucleon system 
\cite{Hall:2014uca}.
The structure of the N*(1535) has also been discussed within
chiral coupled-channel models
\cite{Inoue:2001ip,Hyodo:2008xr,Garzon:2014ida,Kaiser:1995cy}.
Here 
the N*(1535) and N*(1650)
are explained as a $K \Sigma$ 
state together with strong vector meson component
\cite{Garzon:2014ida}.
These coupled-channel model calculations are performed 
with the $\eta$ treated as a pure octet state. 
In Jefferson Laboratory measurements,
the N*(1535) contribution to $\eta$ electroproduction 
was observed to fall away more slowly with increasing 
large $Q^2$ (up to about 7~GeV$^2$)
than expected for a meson-baryon bound system 
\cite{Armstrong:1998wg,Dalton:2008aa,Aznauryan:2011qj,Burkert:2018oyl}. 
This suggests a significant 3-quark contribution.
On the other hand, the low $Q^2$ (below 1 GeV$^2$) 
longitudinal transition amplitude suggests the need 
for meson cloud or other $4 q \overline{q}$ contributions
to the N*(1535) wavefunction.

The branching ratios for the N*(1535) to decay to $\eta$-nucleon 
and pion-nucleon final states are approximately equal, 
about 45\%.
This result is interpreted in \textcite{Olbrich:2017fsd} 
as evidence for a possible gluon anomaly contribution to
the decay.
The strong $\eta$ coupling has also been interpreted 
in quark models with configuration mixing between the
N*(1535) and N*(1650)
\cite{Saghai:2001yd,Chiang:2002ah}.

\section{The $\eta$ and $\eta'$ in nuclei}

There is presently vigorous experimental and theoretical 
activity aimed at understanding 
the $\eta$ and $\eta'$ in medium and 
to search for evidence of possible
$\eta$ and $\eta'$ bound states in nuclei.
Medium modifications need to be understood 
self-consistently within the interplay of confinement,
spontaneous chiral symmetry breaking and axial U(1) dynamics.
In the limit of chiral restoration the pion decay constant 
$f_{\pi}$ should go to zero and 
(perhaps) with scalar confinement 
the pion constituent-quark and pion nucleon coupling 
constants should vanish with dissolution of the pion wavefunction.

One finds a small pion mass shift of order a few MeV in nuclear matter 
\cite{Kienle:2004hq}.
Experiments with deeply bound 
pionic atoms reveal a reduction in 
the value of the pion decay constant
$f_{\pi}^{*2}/f_{\pi}^2 = 0.64 \pm 0.06$ 
at nuclear matter density \cite{Suzuki:2002ae}.
Kaons are observed to experience an effective mass drop 
for the $K^-$ to about 270 MeV at two times 
nuclear matter density in heavy-ion collisions 
\cite{Schroter:1994ck,Barth:1997mk}.
These heavy-ion experiments also suggest the effective 
mass of antiprotons is reduced by about 100-150 MeV 
below their mass in free space \cite{Schroter:1994ck}.
What should we expect for the $\eta$ and $\eta'$ ?
How does the gluonic part of their mass change in nuclei?

Meson masses in nuclei are determined from the meson nucleus 
optical potential and the scalar induced contribution 
to the meson propagator evaluated at zero three-momentum, 
${\vec k} =0$, in the nuclear medium.
Let $k=(E,{\vec k})$ and $m$ denote the four-momentum and 
mass of the meson in free space. 
Then, one solves the equation
\begin{equation}
k^2 - m^2 = {\tt Re} \ \Pi (E, {\vec k}, \rho)
\end{equation}
for ${\vec k}=0$
where $\Pi$ is the in-medium $s$-wave meson self-energy
and $\rho$ is the nuclear density.
Contributions to the in medium mass come from coupling to 
the scalar $\sigma$ field in the nucleus in mean field approximation, 
nucleon-hole and resonance-hole excitations in the medium.
For ${\vec k}=0$, $k^2 - m^2 \sim 2 m (m^* - m)$ 
where $m^*$ is the effective mass in the medium.
The mass shift $m^*-m$ is the depth or real part of 
the meson nucleus optical potential. The imaginary part of 
the potential measures the width of the meson in the nuclear medium.
The $s$-wave self-energy can be written as 
\cite{Ericson:1988gk}
\begin{equation}
\Pi (E, {\vec k}, \rho) \bigg|_{\{{\vec k}=0\}}
=
- 4 \pi \rho \biggl( { b \over 1 + b \langle {1 \over r} \rangle } \biggr) .
\label{Eq34}
\end{equation}
Here 
$
b = a ( 1 + {m \over M} )
$
where 
$a$ is the meson-nucleon scattering length, 
$M$ is the nucleon mass and
$\langle {1 \over r} \rangle$ is the inverse correlation length,
$\langle {1 \over r} \rangle \simeq m_{\pi}$ 
for nuclear matter density.
Attraction corresponds to positive values of $a$.
The denominator in Eq.~(\ref{Eq34}) is the 
Ericson-Ericson-Lorentz-Lorenz double scattering correction.

Studies involving bound state searches and excitation
functions of mesons in photoproduction from nuclear 
targets give information about the meson nucleus
optical potential.

With a strong attractive interaction there is a chance
to form meson bound states in nuclei \cite{Haider:1986sa}.
If found, these mesic nuclei would be a new state of 
matter bound just by the strong interaction.
They differ from mesonic atoms~\cite{Yamazaki:1996zb}
where, for example, a
$\pi^-$ is trapped in the Coulomb potential of the nucleus
and bound by the electromagnetic interaction~\cite{Toki:1989wq}.

Early experiments with low statistics
using photon~\cite{Pheron:2012aj,Baskov:2012yd}, 
pion~\cite{Chrien:1988gn}, 
proton~\cite{Budzanowski:2008fr} 
or deuteron~\cite{Afanasiev:2011zza,Moskal:2010ee}
beams 
gave hints for possible $\eta$ mesic 
bound states but no clear signal \cite{Metag:2017yuh,Kelkar:2013lwa}.
New COSY searches have focused on possible $\eta$ 
bound states in $^3$He and $^4$He \cite{Adlarson:2013xg,Adlarson:2016dme}.
Eta bound states in helium require a large $\eta-$nucleon 
scattering length 
with real part greater than about 
0.7--1.1~fm  
\cite{Barnea:2017epo,Barnea:2017-NP,Fix:2017ani}.
At J-PARC the search for $\eta$-mesic nuclei 
is planned using pion induced reactions on $^7$Li and $^{12}$C targets
\cite{Fujioka:2010dv}.
Recent measurements of $\eta'$ photoproduction from 
nuclear targets have been interpreted 
to mean a small $\eta'$ width in nuclei
$20 \pm 5.0$ MeV at nuclear matter density $\rho_0$ 
\cite{Nanova:2012vw}
that might give rise to relatively narrow bound 
$\eta'$-nucleus states accessible to experiments.
New experimental groups are looking for possible $\eta'$
bound states in carbon using the (p, d) reaction
at
GSI/FAIR \cite{Tanaka:2016bcp,Tanaka:2017cme},
and photoproduction studies at 
Spring-8 with carbon and copper \cite{Shimizu:2017kua}.
Exciting possibilities could also be explored 
at ELSA in Bonn \cite{Metag:2015lza}.
For clean observation of a bound state one needs 
larger attraction than absorption and thus the real 
part of the meson-nucleus optical potential to be much 
bigger than the imaginary part.

\subsection{The $\eta'$ in medium}

The $\eta'$-nucleus optical potential has been measured 
by the CBELSA/TAPS Collaboration in Bonn 
through studies of excitation 
functions in photoproduction experiments from nuclear targets. 
In photoproduction experiments the production cross section 
is enhanced with the lower effective meson mass in the nuclear medium. 
When the meson leaves the nucleus it returns on-shell 
to its free mass with the energy budget conserved at
the expense of the kinetic energy so that excitation functions
and momentum distributions can provide essential clues to the
meson properties in medium \cite{Metag:2011ji,Weil:2012qh}.

Using this physics a first (indirect) estimate of the $\eta'$ 
mass shift has recently been deduced 
by the CBELSA/TAPS Collaboration \cite{Nanova:2013fxl}.
The $\eta'$-nucleus optical potential 
$V_{\rm opt} = V_{\rm real} + iW$
deduced from these photoproduction experiments with a 
carbon target is 
\begin{eqnarray}
V_{\rm real} (\rho_0)
= m^* - m 
&=& -37 \pm 10 \pm 10 \ {\rm MeV}
\nonumber \\ 
W(\rho_0) &=& -10 \pm 2.5 \ {\rm MeV}
\label{Eq:potential}
\end{eqnarray}
at nuclear matter density $\rho_0$.
In this experiment the average momentum of the produced 
$\eta'$ was 1.1 GeV.
The experiment was repeated with a niobium target with
results
$V_{\rm real} (\rho_0)
= -41 \pm 10 \pm 15 \ {\rm MeV}$
and
$ W(\rho_0) = -13 \pm 3 \pm 3 \ {\rm MeV} $
\cite{Friedrich:2016cms,Nanova:2016cyn}.
This optical potential corresponds to an effective
scattering length in medium with real part about 0.5~fm 
in mean field approximation
(switching off the Ericson-Ericson rescattering 
denominator in Eq.~(42)),
consistent with the COSY-11 and photoproduction values,
Eqs.~(\ref{Eq:C11-a-eta-prime},39).
These numbers with small width in medium
suggest that bound states may be within reach of forthcoming experiments.
\begin{figure}
\centering
\includegraphics[width=0.44\textwidth]{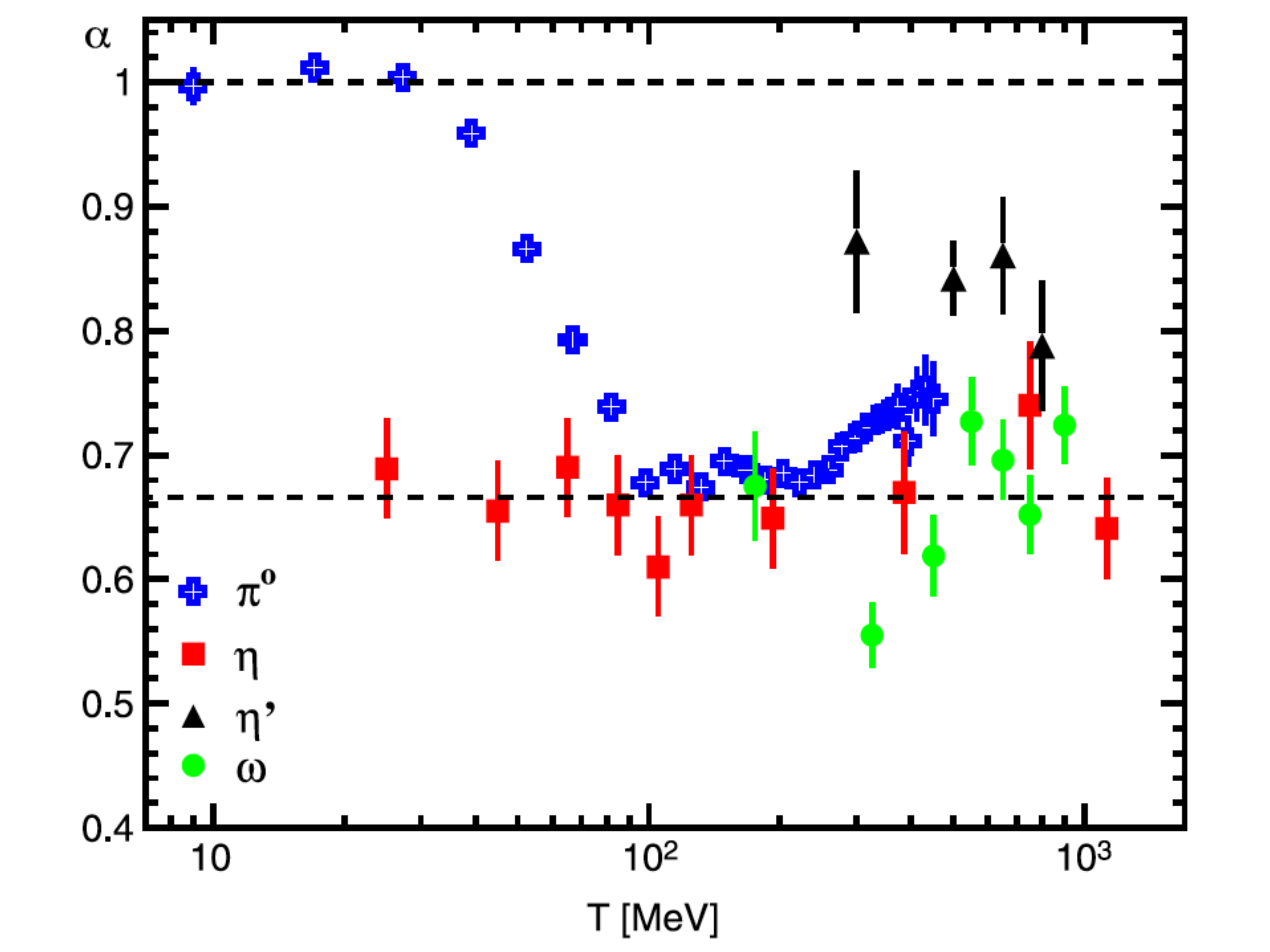}
\caption{
Dependence of the parameter $\alpha$ (Eq.~(\ref{eq:alpha})) on the kinetic energy $T$ of the mesons 
for $\pi^0, \eta, \omega$ and $\eta^\prime$. 
The Figure is taken from \textcite{Nanova:2012vw}.
\label{fig:sigma_abs}
}
\end{figure}

The transparency of nuclei to propagating mesons is
illustrated through Fig.~\ref{fig:sigma_abs}.
Here the cross sections for meson production are
parametrized by
\begin{equation}
 \sigma (A) = \sigma_0 A^{\alpha(T)}
\label{eq:alpha}
\end{equation}
where $\sigma_0$ is the photoproduction cross-section
from a free nucleon and $\alpha$ is a parameter depending
on the meson and its kinetic energy.
The value $\alpha \approx 1$
implies no absorption while 
$\alpha \approx \frac{2}{3}$ corresponds to the meson
being emitted only from the nuclear surface and thus 
strong absorption inside the nucleus.
Fig.~6 shows that
the nucleus is approximately transparent to low-energy 
pions up to the threshold for 
$\Delta$ resonance excitation 
when
$\alpha$ drops to around 2/3, rising slightly at higher
energies.
The $\eta$ and $\omega$ mesons have strong absorption.
For the $\eta'$ one finds $\alpha \approx 0.84 \pm 0.03$
averaged over all kinetic energies
signifying weaker interaction with the nucleus.

The mass shift, Eq.~(\ref{Eq:potential}), 
is very similar to the 
expectations of the Quark Meson Coupling model, QMC~\cite{Bass:2005hn}.
In the QMC model medium modifications are calculated at
the quark level through coupling of the light quarks in 
the hadron to the scalar isoscalar $\sigma$ 
(and also $\omega$ and $\rho$) mean fields in the nucleus,
for a review see 
\textcite{Guichon:1987jp},
\textcite{Guichon:1995ue} and \textcite{Saito:2005rv}.
One works in mean field approximation.
The coupling constants for the coupling of light-quarks 
to the $\sigma$ (and $\omega$ and $\rho$) mean fields in 
the nucleus are adjusted to fit the saturation energy and 
density of symmetric nuclear matter and the bulk symmetry energy. 
The large $\eta$ and $\eta'$ masses are used to motivate 
taking a MIT Bag description for the meson wavefunctions,
\cite{Tsushima:1998qw,Tsushima:1998qp}.
Phenomenologically, the MIT Bag gives a good fit
 to meson properties in free space for the kaons 
 and heavier hadrons \cite{DeGrand:1975cf}.
Gluonic topological effects are understood to be 
``frozen in'', meaning that they are only present 
implicitly through the masses and mixing angle in the model.
The strange-quark component of the wavefunction does not 
couple to the $\sigma$ mean field and $\eta$-$\eta'$ mixing 
is readily built into the model.
Possible binding energies and the in-medium masses of 
the $\eta$ and $\eta'$ are sensitive to the flavor-singlet component in the mesons 
and hence to the non-perturbative 
glue associated with axial U(1) dynamics~\cite{Bass:2005hn}.
Working with the mixing scheme in Eq.~(\ref{eq11})
with an $\eta$-$\eta'$ mixing angle of $-20^\circ$ 
the QMC prediction for the $\eta'$ mass in medium at nuclear 
matter density is 921 MeV, that is a mass shift of $-37$ MeV. 
This value is in excellent agreement with the mass shift 
$-37 \pm 10 \pm 10$~MeV deduced from photoproduction data,
Eq.~(\ref{Eq:potential}). 
Mixing increases the octet relative to singlet component in
the $\eta'$, reducing the binding through increased strange
quark component in the $\eta'$ wavefunction.
Without the gluonic mass contribution the $\eta'$ 
would be a strange quark state after $\eta$-$\eta'$ mixing.
Within the QMC model there would be no coupling to the 
$\sigma$ mean field and no mass shift so that any observed mass shift 
is induced by glue associated with the QCD axial 
anomaly that generates part of the $\eta'$ mass.
For the $\eta$ meson the potential depth predicted by QMC 
is $\approx -100$~MeV at nuclear matter density with 
-20 degrees mixing. 
For a pure octet $\eta$ the model predicts a mass shift of
$\approx -50$~MeV.
Increasing the flavor-singlet component in the $\eta$ at 
the expense of the octet component gives more attraction, 
more binding and a larger value of the $\eta$-nucleon scattering length, $a_{\eta N}$.

In QMC $\eta$-$\eta'$ mixing with the phenomenological 
mixing angle $-20^\circ$ leads to a factor of 
two increase in the mass-shift and 
in the scattering length obtained in the model
relative to the prediction for a pure octet $\eta_8$
\cite{Bass:2005hn}.
This result may explain why values of $a_{\eta N}$ 
extracted from phenomenological fits to experimental 
data where the $\eta$-$\eta'$ mixing angle is unconstrained 
give larger values (with real part about 0.9 fm)
than those predicted 
in theoretical coupled-channel models where the $\eta$ 
is treated as a pure octet state; 
see Section V.

Recent coupled-channel model calculations have appeared 
with mixing and vector meson channels included, 
with predictions for $\eta'$ bound states 
for a range of possible values of $a_{\eta' N}$ 
\cite{Nagahiro:2011fi}.
Larger mass shifts, downwards by up to 80-150 MeV, 
were found in Nambu-Jona-Lasinio model calculations 
(without confinement) 
\cite{Nagahiro:2006dr} 
and in linear sigma model calculations 
(in a hadronic basis) 
\cite{Sakai:2013nba}
which also gave a rising $\eta$ effective mass at finite density.
Different QCD inspired models of the $\eta$ and $\eta'$ 
nucleus systems are constructed with different selections 
of ``good physics input'': 
how they treat confinement, chiral symmetry and axial U(1) dynamics. 
These different theoretical results raise interesting 
questions about the role of confinement and how massive
light pseudoscalar states can be for their wavefunctions 
to be treated as pure Goldstone bosons in the models.

\begin{figure}[t!]
\centering \includegraphics[width=80.0mm]{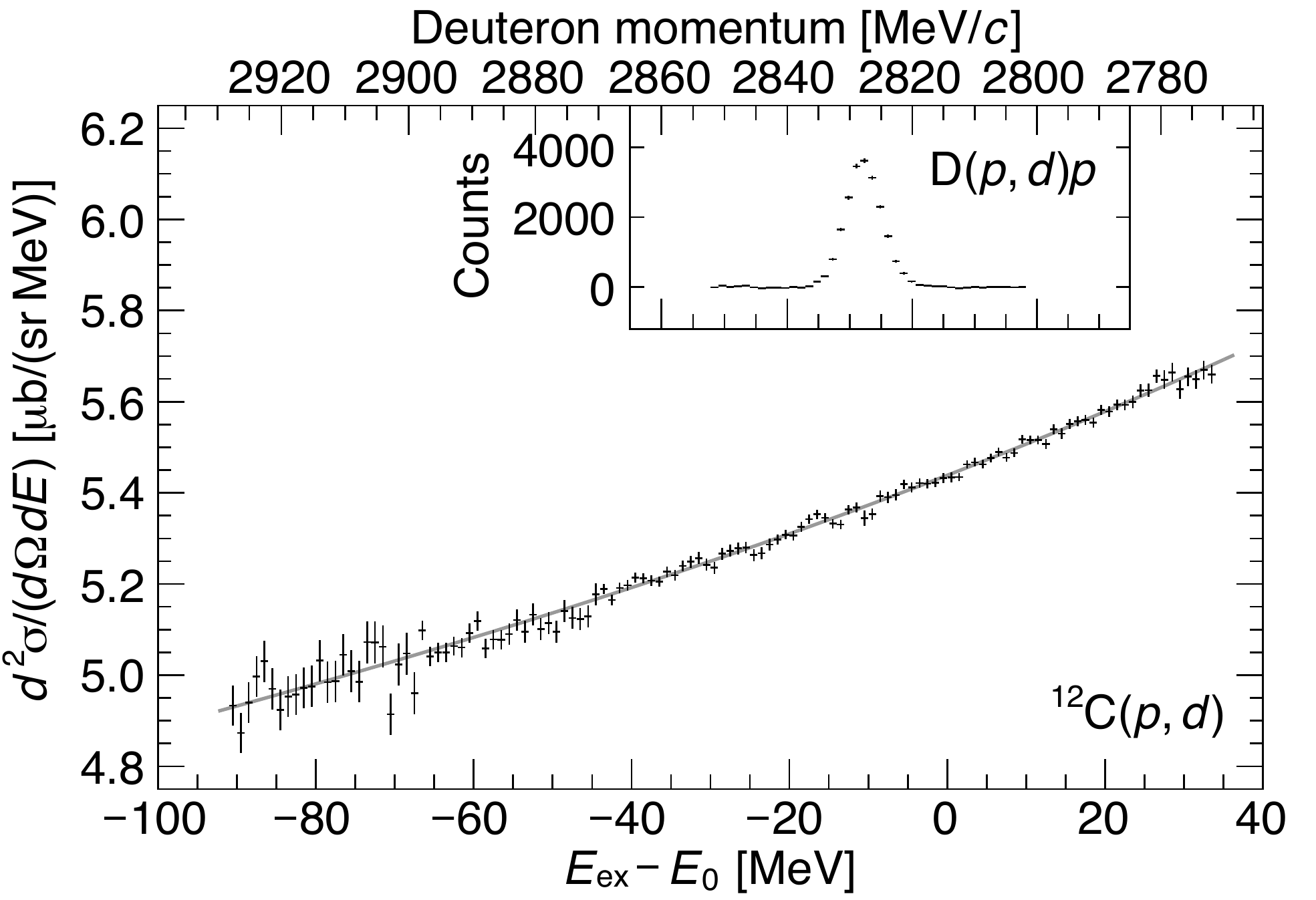}
\caption{\label{fig_spectrum} 
Excitation spectrum of $^{11}$C measured in 
the $^{12}$C($p$,$d$) reaction at a proton energy of 2.5 GeV.
The horizontal axis is the excitation energy 
$E_{\mathrm{ex}}$ referring to the $\eta^{\prime}$ emission threshold $E_0 =  957.78$~MeV. The gray solid curve 
displays a fit with a third-order polynomial.
The inset displays a momentum spectrum of the deuterons 
in the calibration D$(p,d)p$ reaction at 1.6~GeV.
The Figure is taken from 
\textcite{Tanaka:2017lxw,Tanaka:2016bcp}.
} 
\end{figure}

Experiments in heavy-ion collisions \cite{Averbeck:1997ma} 
and $\eta$ photoproduction from nuclei 
\cite{RoebigLandau:1996xa,Yorita:2000bu} 
suggest little modification of the N*(1535) 
excitation in-medium, 
though some evidence for the broadening of the 
N*(1535) 
in nuclei was reported in \textcite{Yorita:2000bu}.
In the QMC model
the excitation energy is $\sim 1544$ MeV, 
consistent with observations,
with the scalar attraction compensated by repulsion 
from coupling to the $\omega$ mean field~\cite{Bass:2005hn}.
The QMC model predictions for the kaon and proton mass 
shifts are a reduction in the $K^-$ mass of about 
100 MeV and effective proton mass about 
755 MeV at nuclear matter density \cite{Saito:2005rv}.

The first experiments to search for possible $\eta'$
bound states in carbon have been performed at GSI
with inclusive measurement of the 
$^{12}\hspace{-0.03cm}\mbox{C}$(p,d) 
reaction~\cite{Tanaka:2016bcp,Tanaka:2017cme};
see Fig.~7.
These experiments exclude very deeply bound narrow states 
corresponding to real part of the optical potential larger than about 150 MeV 
predicted~\cite{Nagahiro:2012aq,Nagahiro:2006dr} 
based on the NJL model 
when assuming the $\eta'$ absorption 
(imaginary part of the potential of -10~MeV) 
deduced 
from measurements of the transparency in nuclei, 
Eq.~(44)~\cite{Nanova:2012vw,Friedrich:2016cms}.
More precise studies are planned using 
semi-inclusive and exclusive measurements 
with the registration of the decay products of the mesic state
\cite{Tanaka:2017lxw}.

\subsection{$\eta$ mesic nuclei}

\begin{figure}[b!]
\centering
  \includegraphics[width=0.44\textwidth]{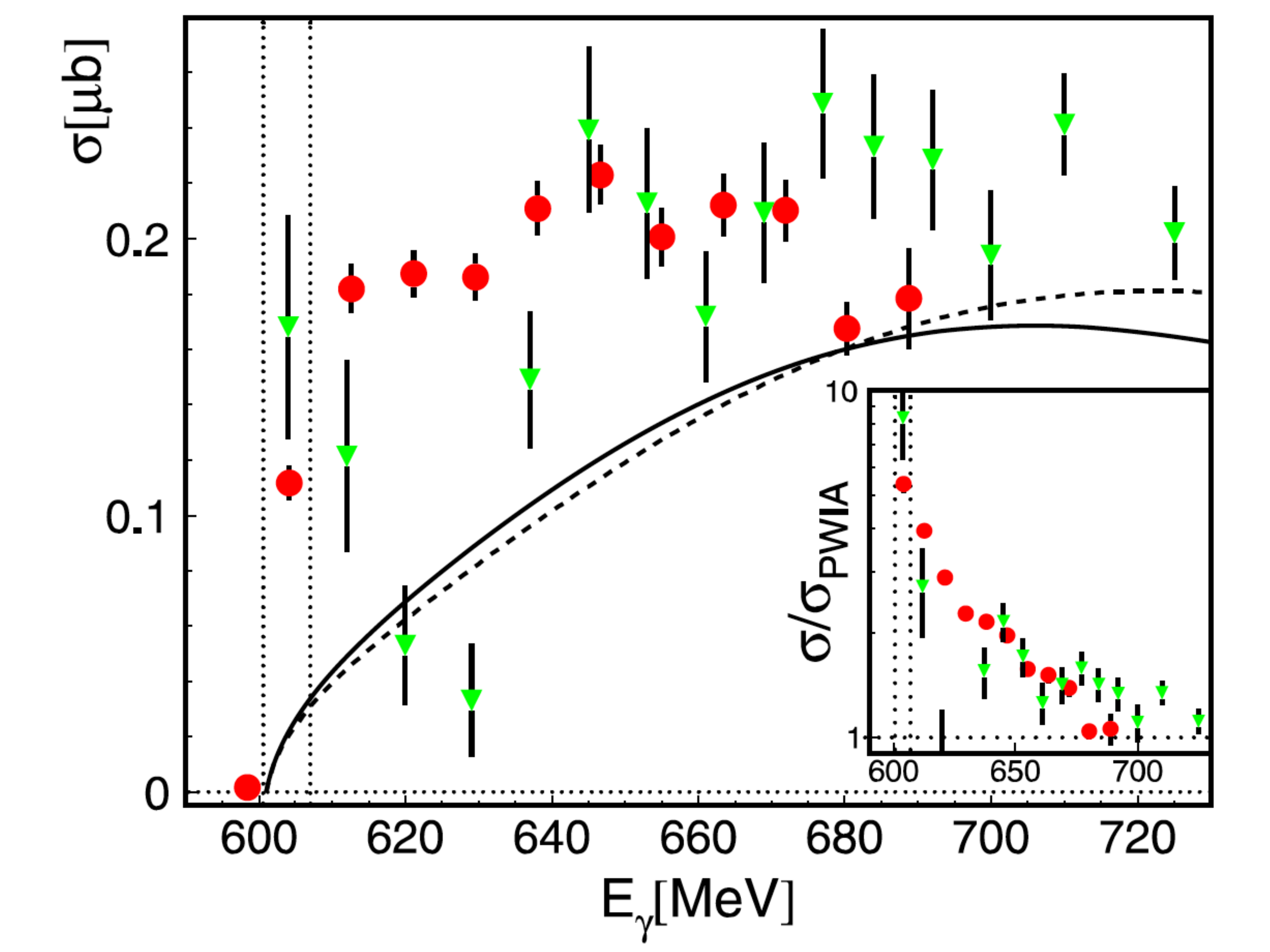}
\caption{
Total cross section for the $\gamma ^3$He$ \rightarrow \eta ^3$He reaction. 
Data are from references \cite{Pheron:2012aj} (red points) and \cite{Pfeiffer:2003zd} (green triangles).
Solid (dashed) curves represent plane wave impulse approximation (PWIA) 
calculations with a realistic (isotropic) angular distribution for the $\gamma n \rightarrow n \eta$ reaction. 
Insert: ratio of measured and PWIA cross sections. 
The Figure is taken from 
\textcite{Pheron:2012aj}.
\label{fig:3He-photo}
}
\end{figure}

\begin{figure}[t!]
\vspace{-4cm}
        \centerline{
        \includegraphics[width=0.49\textwidth]{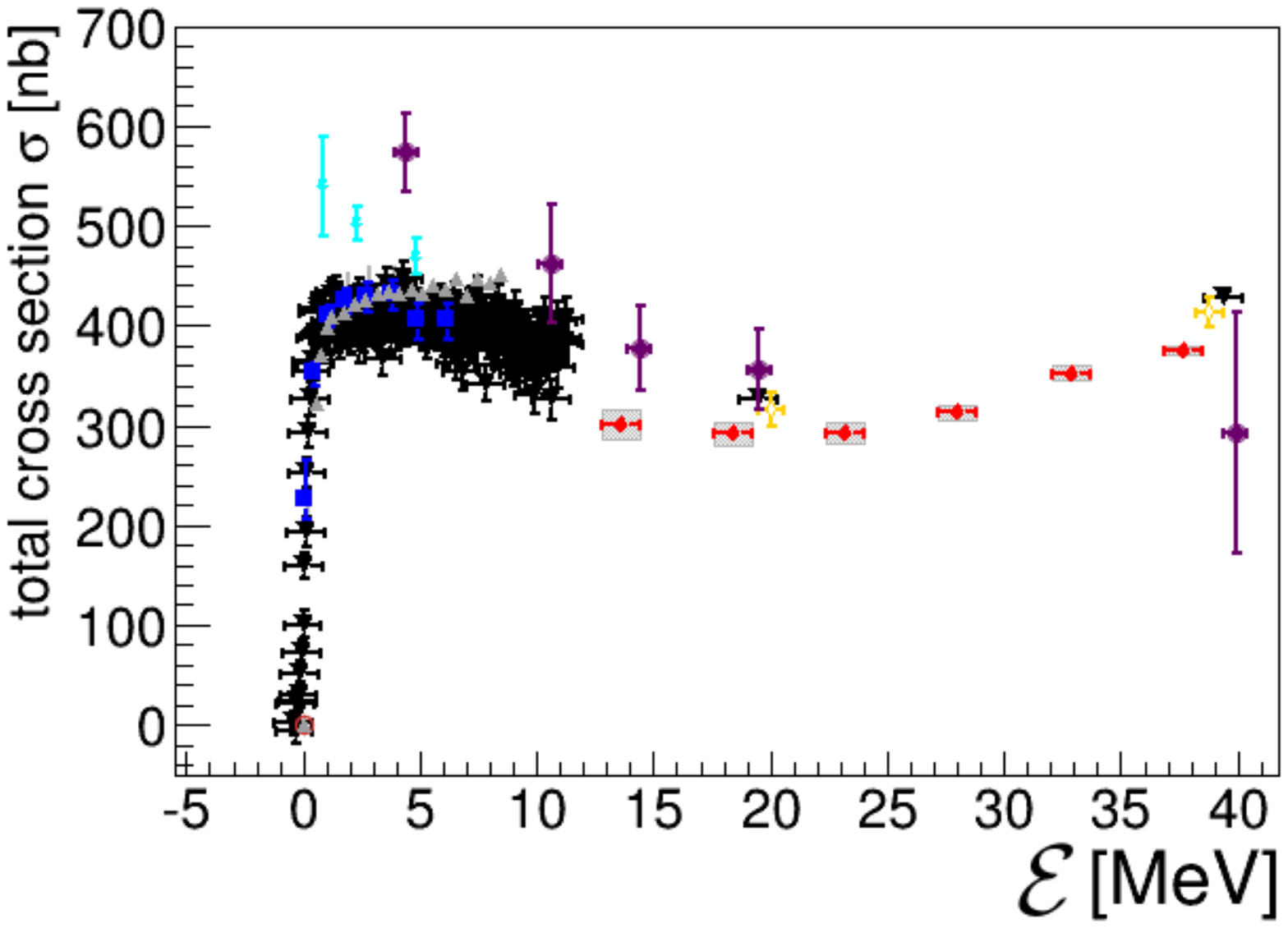}}
\vspace{-3.5cm}
\caption{ 
World data on the $pd\rightarrow {}^3\text{He}\,\eta$
reaction close-to-threshold~\cite{Berger:1988ba,Mayer:1995nu,Betigeri:1999qa,Bilger:2002aw,Smyrski:2007nu,Adam:2007gz,Mersmann:2007gw,Rausmann:2009dn,Adlarson:2014ysb,Adlarson:2018rgs}. 
Notice the sharp rise at threshold.
The Figure is adapted from~\textcite{Adlarson:2018rgs}.
\label{fig:totalcross}
}
\end{figure}

Hints for possible $\eta$ helium bound states are 
inferred from observed strong interaction in the 
$\eta$ helium system.
One finds a sharp rise in the cross section 
at threshold for $\eta$ production in both
photoproduction from $^3$He 
and in the proton-deuteron reaction
$dp \to ^3\!\!He \ \eta$, 
which may hint at a reduced $\eta$ effective mass 
in the nuclear medium.
For these data see 
Fig.~\ref{fig:3He-photo} and
Fig.~\ref{fig:totalcross} respectively.
One also finds a small and constant value of the 
analyzing power 
~\cite{Papenbrock:2014hup} 
as well as strong variation of the angular asymmetry 
for $\eta$ meson emission~\cite{Smyrski:2007nu,Mersmann:2007gw}
indicating strong changes of the phase of the 
$s$-wave production amplitude 
with energy, as expected 
with a bound or virtual ${^{3}\mbox{He}}-\eta$ state~\cite{Wilkin:2007aa}.
Sharp but less steep rise in the cross section is 
also seen in the $dd \to ^4\!\!He \ \eta$ reaction 
\cite{Budzanowski:2008qx,Wronska:2005wk,willis:1997ix,Frascaria:1994va}.

Searches for $\eta$ mesic nuclei are ongoing 
with data from the WASA-at-COSY experiment.
The focus has so far been on the reaction
$dd \to \ ^3He N \pi$,
in particular 
studies of the excitation function around the threshold
for $dd \to \ ^4\!He \eta$.
These excitation functions did not reveal a structure 
that could be interpreted as a narrow mesic nucleus. 
Upper limits for the total cross sections for 
bound state production and decay in the processes
$dd \to (^4He-\eta)_{bound} \to ^3\!\!He \ n \pi^0$ 
and 
$dd \to (^4He-\eta)_{bound} \to ^3\!\!He \ p \pi^-$ 
were determined 
assuming the mesic bound state width lies in the range 
5 -- 50 MeV.
Taking into account recent results on the N*(1535) 
momentum distribution in the N*-$^3$He 
nucleus~\cite{Kelkar:2016uwa,Kelkar:2015qbb},
the latest upper limits are about 5~nb and 10~nb for 
the $n \pi^0$ and $p \pi^-$ channels respectively
 \cite{Adlarson:2016dme}. 
These upper limits can be compared to 
model predictions.
For example, within 
the optical model of \textcite{Ikeno:2017xyb}
most of the model parameter space is excluded allowing 
values of the real and imaginary parts of the
potential only between zero and 
about -60~MeV and -7~MeV
respectively~\cite{Skurzok:2018paa}.  
While the achieved experimental sensitivity of 
a few nanobarns is too small to make definite 
conclusions about the existence of a 
$^4$He-$\eta$ bound state, 
the situation with $^3$He may be more positive.
The measurements have similar accuracy of order a few
nanobarns
with
the expected bound state production cross sections 
for $pd\to (^3He-\eta)_{bound}$\cite{Wilkin:2014mla} 
expected to be more than 20 times larger than for
$dd\to (^4He-\eta)_{bound}$~\cite{Wycech:2014wua}.
Data analysis for the pd reaction 
is ongoing~\cite{Rundel:2017rea}.
Recent calculations in the framework of 
optical potential~\cite{Xie:2016zhs},
multi-body calculations~\cite{Barnea:2017-NP},
and pionless effective field theory~\cite{Barnea:2017epo}
suggest a possible $^3$He-$\eta$ bound state.

\subsection{The $\eta'$ at finite temperature}

In addition to finite density, axial U(1) symmetry is
also expected to be (partially) restored at finite 
temperature \cite{Kapusta:1995ww}.
This result is observed 
in recent QCD lattice calculations 
\cite{Bazavov:2012qja,Cossu:2013uua,Tomiya:2016jwr}.
Experimentally, 
there are hints in RHIC data from relativistic 
heavy ion collisions
for a possible $\eta'$ mass suppression at finite 
temperature, 
with claims of at least -200 MeV mass shift
deduced from studies of the intercept $\lambda$
measured in two-charged-pion
Bose-Einstein correlations 
\cite{Csorgo:2009pa,Vertesi:2009wf}. 
With decreasing $\eta'$ mass one expects 
a drop in this parameter at small
transverse momentum~\cite{Vance:1998wd}.
The $\lambda$ parameter 
accounts for the fact that not all pion pairs 
are correlated,
{\it e.g.}, as daughters of long-lived strongly 
decaying resonances and effects from the source dynamics.
A key issue in the analysis here is the matching of this 
dilution factor between experiment and theory.
The ALICE Collaboration at CERN 
see similar effects in the data to the RHIC experiments
with $\lambda$ falling by $\sim 70\%$ at the smallest transverse momentum
without attempting an
$\eta'$ mass shift extraction~\cite{Adam:2015vna}.

\section{High-energy $\eta$ and $\eta'$ production}

In higher energy experiments with proton-proton collisions 
at 450 GeV, or center of mass energy of 28 GeV, 
the WA102 Collaboration at CERN observed that 
central production of $\eta$ and $\eta'$ mesons 
seems to have a similar production mechanism 
which differs 
from that of the $\pi^0$ \cite{Barberis:1998ax}.
This result has been interpreted in terms of 
gluonic pomeron-pomeron
and pomeron-Reggeon fusion
\cite{Close:1999bi,Lebiedowicz:2013ika}.
The pomeron is a non-perturbative color-singlet combination of 
gluon exchange which governs the high energy behavior of hadron 
scattering processes. 
Reggeons involve the sum over meson-like exchanges carrying
particular quantum numbers in these reactions
\cite{Landshoff:1994up,Collins:1984tj}.

Semi-inclusive $\eta$ production in high-energy collisions has been a 
topical issue since the pioneering work of \textcite{Field:1976ve}.
One finds the interesting result that the ratio of $\eta$ 
to $\pi^0$ production rises rapidly with the transverse momentum $p_t$ 
of the produced meson and levels off at at 
$R_{\eta/\pi^0} \sim 0.4 - 0.5$ 
above 
$p_t \sim 2$ GeV
in nuclear collisions 
(proton-proton, proton-nucleus and nucleus-nucleus) 
independent of the colliding nuclei;
see Fig.~10.
These results hold over a wide range of center-of-mass 
energy
($\sqrt{s_{NN}} \sim 30-8000$ GeV)
as well as 
meson production carrying momentum fraction
$x_p > 0.35$ 
of the exchanged photon in
electron positron collisions at LEP, $\sqrt{s} = 91.2$ GeV.

\begin{figure}[t!]
\vspace{-4cm}
\centering
  \includegraphics[width=0.43\textwidth]{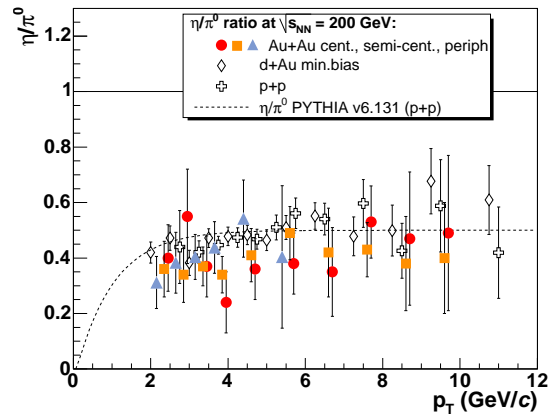}
\caption{
Ratio of $\eta$ to $\pi^0$ production in RHIC PHENIX data.
The Figure is taken from \textcite{Adler:2006hu}.
\label{fig:RHICetapi}
}
\end{figure}

In these relativistic heavy-ion collisions
the invariant yields per nucleon-nucleon collision are
increasingly depleted with centrality in comparison to 
proton-proton results at the same center-of-mass energy.
The maximum suppression factor is about 5 in central Au+Au
collisions~\cite{Adler:2006hu}.
The measured $\eta/\pi^0$ ratio 
is independent of both 
the reaction centrality as well as
the species of colliding protons or nuclei.
These results indicate that any initial and/or final state 
nuclear effects influence the production of light neutral 
mesons at large $p_t$ in the same way.
The approximately constant ratio for $\eta$ to $\pi^0$
production
indicates that the parent quark or gluon parton
first loses energy in the dense medium of the collision
and then fragments into leading mesons $\eta$ and $\pi^0$ 
in the vacuum according to the same probabilities that govern 
high $p_t$ hadron production in more elementary $e^+ e^-$ and
proton-proton collisions.
These results observed at RHIC in 
PHENIX \cite{Adler:2006bv,Adler:2006hu}
and STAR \cite{Abelev:2009hx} data at 
$\sqrt{s_{NN}}=200$ GeV
are also observed by ALICE at the LHC up to 8 TeV
\cite{Acharya:2018hzf,Acharya:2017tlv,Acharya:2018yhg},
with earlier measurements summarized in 
\textcite{Adler:2006bv}.

The fragmentation functions for $\eta$ production in 
high energy processes are discussed in 
\textcite{Aidala:2010bn}.
First measurements of $\eta'$ production in proton-proton
collisions at center of mass energy 200 GeV are reported
by the PHENIX Collaboration in \textcite{Adare:2010fe}.
In ALEPH data from LEP
$\eta'$ production was observed to be anomalously suppressed 
compared to the expectations of string fragmentation models 
without an additional ``$\eta'$ suppression factor'', 
possibly associated with the mass of the produced $\eta'$
\cite{Barate:1999gb}.
The cross section and double helicity asymmetry for $\eta$
production is studied by PHENIX at midrapidity
with comparison to $\pi^0$ production in \textcite{Adare:2010cy}.
The transverse single-spin asymmetry for forward $\eta$ 
production looks as large as if not larger than that 
for forward $\pi^0$ production 
-- see 
PHENIX \cite{Adare:2014qzo} and STAR \cite{Adamczyk:2012xd}
-- and may be related to quark-gluon correlation functions.

\subsection{$\eta'$--$\pi$ interactions and $1^{-+}$ exotics}

\begin{figure}[t!]
\centering
  \includegraphics[width=0.43\textwidth]{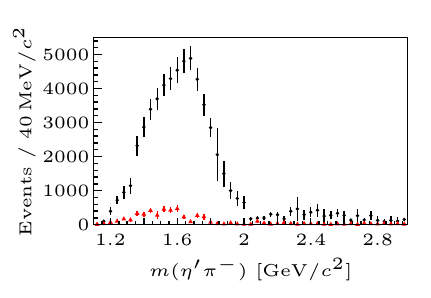}
\caption{
The $\eta' \pi^-$ exotic partial wave 
$1^{-+}$ 
(upper data)
is much enhanced compared to $\eta \pi^-$
(lower data)
in exclusive production from 191 GeV $\pi^-$ scattering
on a hydrogen fixed target.
The Figure is taken from \textcite{Adolph:2014rpp}.
\label{fig:Compass-exotic}
}
\end{figure}

Following the discussion in Section II, 
the OZI-violating interaction
$\xi Q^2 \partial_{\mu} \pi_a \partial^{\mu} \pi_a$
gives a potentially 
important tree-level contribution to the decay 
$\eta' \rightarrow \eta \pi \pi$ \cite{DiVecchia:1980vpx}.
Suppose one takes $\xi$ as negative with attractive interaction.
When iterated in the Bethe-Salpeter equation for $\eta' \pi$
rescattering this interaction 
then yields a dynamically generated 
resonance with quantum numbers $J^{PC} = 1^{-+}$ and 
mass about 1400~MeV.
The dynamics here is mediated by the singlet 
OZI-violating coupling
of the $\eta'$ \cite{Bass:2001zs}.
One finds a possible dynamical interpretation of 
light-mass $1^{-+}$ exotic states,
{\it e.g.}, as observed in experiments at BNL 
\cite{Thompson:1997bs,Chung:1999we,Adams:1998ff,Ivanov:2001rv}
and CERN \cite{Abele:1998gn},
see also \textcite{Szczepaniak:2003vg}.
This OZI-violating interaction will also contribute to 
higher $L$ odd partial waves with quantum numbers $L^{-+}$.
These states are particularly interesting because 
the quantum numbers $1^{-+}, 3^{-+}, 5^{-+}$...
are inconsistent with a simple quark-antiquark bound state.
The COMPASS experiment at CERN has recently measured 
exclusive production of $\eta' \pi^-$ and $\eta \pi^-$ 
in 191 GeV $\pi^-$ collisions on a hydrogen target
\cite{Adolph:2014rpp}.
They find the interesting result that $\eta' \pi^-$ 
production is enhanced relative to $\eta \pi^-$ production 
by a factor of 5-10 in the exotic $L=1,3,5$ partial waves 
with quantum numbers $L^{-+}$ in the inspected invariant 
mass range up to 3 GeV; see Fig.~11.
No enhancement was observed in the even $L$ partial waves.
For further recent discussion, 
see also \textcite{Rodas:2018owy}.

Glueballs, postulated bound states of gluons with integer spin, 
may also couple strongly to the $\eta'$ and $\eta$.
Glueball states are found in lattice pure glue theory
with mixing with quark-antiquark mesons induced in full QCD
\cite{Gui:2012gx,Morningstar:1999rf,Gregory:2012hu,Sun:2017ipk}.
The lightest glueball state is expected to be a scalar 
with the prime candidates discussed in the literature
being the $f_0 (1500)$ and $f_0 (1710)$ states,
much heavier than the lightest mass 
quark-antiquark state -- the pseudoscalar pion.
We refer to 
\textcite{Frere:2015xxa} and \textcite{Brunner:2015oga}
for recent discussion of scalar glueball decays to 
$\eta$ and $\eta'$ final states.
Particularly interesting is a pseudoscalar glueball
in the mass range 2-3 GeV where recent calculations 
suggest a narrow state and very restricted decay pattern
involving $\eta$ or $\eta'$ mesons 
that can be searched for in central exclusive production experiments, 
{\it e.g.}, at the LHC 
\cite{Brunner:2016ygk}.

\section{Summary and Future Challenges}

The isoscalar $\eta$ and $\eta'$ mesons are sensitive 
to the interface of chiral and non-perturbative dynamics.
One finds a rich phenomenology involving 
OZI-violation,
meson production dynamics 
from threshold through to high-energy collisions
and the coupling to new excited nucleon resonances.
Axial U(1) symmetry is expected 
to be partially restored
in QCD media at finite densities and temperature.
This, in turn, leads to predictions 
for the $\eta$ and $\eta'$
effective mass shifts in medium 
and possible meson bound states in nuclei.
The non-perturbative glue 
which generates the large $\eta$ and $\eta'$ masses
also has the potential to induce strong CP violation
in the neutron electric dipole moment which is not observed.
A possible solution to this strong CP puzzle
is connected with a new axion particle which, 
if it exists, might also be associated with dark matter.
Understanding the $\eta$ and $\eta'$ systems 
is important to nuclear, high-energy and astrophysics.

New experiments will give valuable insight into
$\eta$ and $\eta'$ physics.
The search for $\eta$ and $\eta'$ mesic nuclei 
will help pin down the dynamics of axial U(1) symmetry
breaking in low-energy QCD.
Determining the $\eta'$ properties 
at finite temperature 
in relativistic heavy-ion collisions would further probe 
axial U(1) dynamics in the QCD phase diagram.
Precision studies of $\eta$ and $\eta'$ decays 
are a probe for new physics beyond the Standard Model.
Production of $\eta'$ mesons in connection with glueball
production will test theoretical ideas about gluonic
excitations in non-perturbative QCD.

\section*{Acknowledgments}

We thank 
C.~Aidala,
V.~Burkert,
M.~Faessler,
A.~Fix,
S.~Hirenzaki,
K.~Itahashi, 
J.~Krzysiak,
W.~Melnitchouk,
V.~Metag, 
G.~Moskal, 
E.~Oset,
M.~Pitschmann,
A.~Rebhan,
H.~Shimizu,
M.~Silarski,
M.~Skurzok, 
Y.~Tanaka,
A.~W.~Thomas and 
M.~Weber for helpful discussions.
We acknowledge support from the Polish National Science 
Centre through the grant No. 2016/23/B/ST2/00784
and the Foundation for Polish Science through the TEAM/2017-4/39 programme.

\bibliography{rmp3}

\end{document}